\documentclass[final, 5p, times, twocolumn]{elsarticle}

\usepackage{lineno}  

\usepackage{graphicx}%
\usepackage{siunitx}
\usepackage{bm}
\usepackage[caption=false]{subfig}
\usepackage{mathtools}
\usepackage{multirow}
\usepackage{amssymb}
\usepackage{braket}
\usepackage{booktabs}
\usepackage{placeins}
\usepackage{cancel}
\usepackage{xcolor}
\usepackage{spreadtab}
\usepackage{dsfont}
\usepackage[version=4]{mhchem}
\usepackage[colorlinks=true, urlcolor=cyan, citecolor=blue, linkcolor=blue]{hyperref}
\usepackage{float}
\usepackage{tabularx}
\usepackage{ragged2e}

\usepackage{array}

\newcolumntype{L}{>{\hsize=1.2\hsize}X}   %
\newcolumntype{S}{>{\hsize=0.925\hsize}X} %
\DeclareUnicodeCharacter{2009}{\,}

\newcommand\absq{\ensuremath{\mathit{q}}}
\newcommand{\Tc}{\ensuremath{T_{\mathrm c}}}

\newcommand{\GSFE}{\ensuremath{\gamma^{\mathrm{GSFE}}}}
\newcommand{\USFE}{\ensuremath{\gamma^{\mathrm{USFE}}}}
\newcommand{\ISFE}{\ensuremath{\gamma^{\mathrm{ISFE}}}}
\newcommand{\UTE}{\ensuremath{\gamma^{\mathrm{UTE}}}}

\usepackage{xr}
\begin{document}

\title{In search of novel ductile superconductors}

\author[1]{Yiming Zhang}
\author[1,2]{Samuel Ponc\'e\corref{cor1}}
\ead{samuel.ponce@uclouvain.be}

\cortext[cor1]{Corresponding author}

\address[1]{%
European Theoretical Spectroscopy Facility, Institute of Condensed Matter and Nanosciences, Université catholique de Louvain, Chemin des Étoiles 8, B-1348 Louvain-la-Neuve, Belgium. 	
}
\address[2]{%
WEL Research Institute, avenue Pasteur 6, 1300 Wavre, Belgium.		
}%
 

\date{\today}

\begin{abstract}
We performed a first-principles high-throughput screening of the mechanical properties of phonon-mediated superconductors selected from the recent experimentally synthesized superconducting materials database [\href{https://journals.aps.org/prxenergy/abstract/10.1103/sb28-fjc9}{PRX Energy \textbf{4}, 033012 (2025)}].
We developed the workflows that combine first-principles calculations of elastic constants and generalized stacking fault energies to assess the ductility of superconducting candidates.
Starting from the 250 materials identified with promising superconducting critical temperatures, we computed their elastic tensors to evaluate bulk and shear moduli, Pugh's ratio, and Pettifor's ratio from first principles.
To further characterize their plastic deformation behavior, we calculated the stacking fault energy and surface energy for selected materials and slip directions, allowing the estimation of Rice's ratio and ductility indicators.
We found that several new materials simultaneously exhibit high-\Tc~and ductility including \ce{HfPd2Al}, \ce{TiRuSb}, and \ce{ZrNi2Ga} with predicted isotropic \Tc= 6.80~K, 12.88~K, and 8.23~K, respectively.
This work offers a quantitative mapping of mechanical performance across a wide range of superconductors and provides a reference to identify new mechanically promising superconductors.
\end{abstract}

\maketitle

\section{Introduction}

Superconductors play an important role in a wide range of technological applications including medical imaging, particle accelerators, energy transmission, magnetic levitation, and quantum computing~\cite{Bray2009,Scalapino2012,Chen2024}. 
However, many practical implementations also require superconductors to exhibit excellent mechanical properties such as high strength and ductility~\cite{Shi2024}.
Currently, widely used superconductors are often limited by their mechanical performance and have not yet reached sufficiently high superconducting transition temperatures~(\Tc), even within the framework of conventional Bardeen-Cooper-Schrieffer~(BCS) theory~\cite{Bardeen1957}.
Recently, the \textsc{supercond-EPW}~\cite{Bercx2025}  database reports theoretical predictions for numerous new BCS superconductors using the state-of-the-art anisotropic Migdal-Eliashberg formalism~\cite{Margine2013,Mori2024}. 
Crucially, the database contains only experimentally synthesized materials, which ensures the reliability of the superconducting prediction.
To assess the practical viability of these candidates for industrial applications, it is crucial to evaluate their mechanical properties alongside their superconducting transition temperatures \Tc.

Within the linear elastic regime, the mechanical behavior can be characterized through the stress-strain response, which can be obtained using first-principles calculations based on density functional theory~(DFT). 
Several high-throughput studies have reported elastic tensors for various material families, including MXenes~\cite{Tian2022}, halide perovskites~\cite{Diao2022}, and ceramics~\cite{Xiao2020}. 
Mechanical properties such as Pugh's ratio~\cite{Pugh1954}, Pettifor's criterion~\cite{Pettifor1992}, and Poisson's ratio have been used to differentiate brittle from ductile behavior. 
These metrics can be computed from first principles and thus lend themselves well to high-throughput analysis. 
Although Pugh's ratio is reliable for elemental metals, it may not perform well in complex compounds, as suggested by studies on C15-type materials~\cite{Long2016}. 
To improve ductility assessment, Kelly \textit{et al.}~\cite{Kelly1967} proposed a method that considers the ratio of maximum shear to bulk modulus, with the aim of describing crack initiation under loading conditions. 
The ideal shear strength denotes the theoretical stress necessary to initiate plastic deformation in a defect-free crystal and thus provides an upper limit for real materials. 
However, plasticity is mediated by dislocations that provide a lower energy pathway for deformation compared to ideal shearing.
Consequently, the experimentally measured shear strengths are markedly lower than their ideal counterparts.
Dislocations are carriers of plasticity and have been examined from first principles in elemental metals~\cite{Yan2004}, which provides insights into the deformation mechanisms on the atomic scale.
Generalized stacking fault energies~(GSFE) provide key insights into the behavior of dislocations in solids~\cite{Woodward2008,Henager2005,Vitek2004}.
High-throughput studies of GSFE have been performed for various material classes, including elemental metals~\cite{Tu2019, Rodney2017, Lu2001}, alloys~\cite{Vamsi2021, Siegel2005}, two-dimensional van der Waals materials~\cite{Gao2022}, and ceramics~\cite{Zhang2024, Koutn2021}.
To assess the propensity for dislocation emission versus cleavage, Rice and Thomson introduced a ductility indicator based on the ratio of unstable stacking fault energy~(USFE) to surface energy, known as Rice's ratio~\cite{Rice1974}.
This metric provides an indirect measure of the energetic cost associated with crack propagation versus plastic slip.

In this work, we investigate the mechanical properties of superconducting materials from the \textsc{supercond-EPW} database. 
We begin by comparing elastic moduli obtained via the Born expansion~\cite{Lin2026} and the finite-displacement~(FD) method on the unit cell~\cite{DalCorso2015,DalCorso2016}, verifying that the Born expansion yields accurate elastic constants when the phonon grid is sufficiently converged. 
Subsequently, we compute the Pugh's and Pettifor's ratios for 250 metals, for which the isotropic superconducting \Tc~is predicted, to find ductile superconducting candidates.
We then compute the GSFE for 19 materials to gain insight into the dislocation behavior and plastic deformation mechanisms of these materials and propose a new ductility indicator. 
Using this new indicator, we find that \ce{HfPd2Al}, \ce{TiRuSb}, and \ce{ZrNi2Ga} are the most promising novel ductile superconductors from the supercond-EPW database.
Our results guide the search for ductile superconductors suitable for practical applications.

\section{Results}

Elastic properties can be computed with finite-displacement~\cite{DalCorso2015,DalCorso2016} or density functional perturbation theory~(DFPT)~\cite{Gonze1997,Baroni2001}.
Most DFPT codes give access to interatomic force constants~(IFC).
Using the Born expansion~\cite{Lin2026}, one can reuse the existing high-throughput phonon databases~\cite{Petretto2018,Bercx2025} to compute elastic constants directly from IFCs at negligible computational cost.   
We validate the Born expansion by computing the bulk modulus $B$, shear modulus $G$, and elastic constants $C$ from the Born expansion and from FD.   

We applied the Born expansion~\cite{Lin2026} to the 250 metals of the \textsc{supercond-EPW}~\cite{Bercx2025} database.
The database was created using \textsc{AiiDA}~\cite{Huber2020, Uhrin2021}, \textsc{Quantum ESPRESSO}~\cite{Giannozzi2017}
and the \textsc{EPW} code~\cite{Ponce2016,Lee2023}.
We used the same norm-conserving Perdew–Burke–Ernzerhof~(PBE) pseudopotentials from the scalar-relativistic table of \textsc{PseudoDojo} v0.5~\cite{vanSetten2018}, convergence parameters, and structural parameters.
We computed elastic properties using \textsc{thermo\_pw}~\cite{thermopw}, based on the generalized Hooke's law relating the stress tensor $\sigma$ and the strain tensor $\varepsilon$ using Voigt notation:  
\begin{align}
	\sigma_{i} &= \sum_j C_{ij} \varepsilon_{j},
\end{align}
where we used the Einstein summation convention.
From the elastic constants, the Voigt-Reuss-Hill~(VRH) average of the bulk and shear modulus are given by~\cite{Hill1952}:
\begin{align}
    B =& (B^{\rm V} + B^{\rm R})/2 \\
    G =& (G^{\rm V} + G^{\rm R})/2,
\end{align}
where $B^{\rm V}$ and $G^{\rm V}$ are the upper bound Voigt estimates and $B^{\rm R}$ and $G^{\rm R}$ are the lower bound Reuss estimates, given by:
\begin{align}\label{eq:bulk_shear_modulus}
    9B^{\rm V} =& C_{11} + C_{22} + C_{33} + 2(C_{12}+ C_{13} + C_{23})\\
    15 G^{\rm V} =& C_{11} + C_{22} + C_{33} - (C_{12}+ C_{13} + C_{23}) \nonumber \\
                  & + 3(C_{44}+ C_{55} + C_{66}) \\ 
    (B^{\rm R})^{-1} =& S_{11}+S_{22}+S_{33}+2(S_{12}+S_{13}+S_{23}) \\
    15(G^{\rm R})^{-1} =& 4(S_{11} + S_{22} + S_{33}) - 4(S_{12}+ S_{13} + S_{23}) \nonumber \\
                  & + 3(S_{44}+ S_{55} + S_{66}),
\end{align}
where $S_{ik}$ is the compliance tensor that satisfies $\sum_k S_{ik}C_{kj} = \delta_{ij}$.
From these linear elastic quantities, the Pugh's ratio $r^{\rm Pugh}$, defined as
\begin{align}
    r^{\rm Pugh} \equiv & \frac{G}{B},
\end{align}
quantifies how shear deformation resistance compares to volume-deformation resistance.
For most crystalline materials, the brittle to ductile transition occurs at $r^{\rm Pugh} \approx 0.57$~\cite{Pugh1954, Thompson2018, Niu2012}.
Pugh's ratio captures the balance between bond rigidity and tendency of plastic deformation.
Another important ratio is the Pettifor's ratio $r^{\rm Pett}$ which is applicable to cubic crystals. 
It was originally defined as the ratio of Cauchy's pressure $C'' = C_{12} - C_{44}$ over Young's modulus $E$~\cite{Pettifor1992} and then revised as the ratio of Cauchy's pressure over bulk modulus $B$~\cite{Senkov2021},
\begin{align}
    r^{\rm Pett} \equiv & \frac{C_{12} - C_{44}}{B}.
\end{align}

Materials with positive Pettifor's ratio are considered ductile~\cite{Pettifor1992}. 
Both elastic descriptors are empirical indicators of brittle-ductile tendencies, but neither describes plastic deformation.

\begin{figure}[!b]
    \centering
    \includegraphics[width=0.99\linewidth]{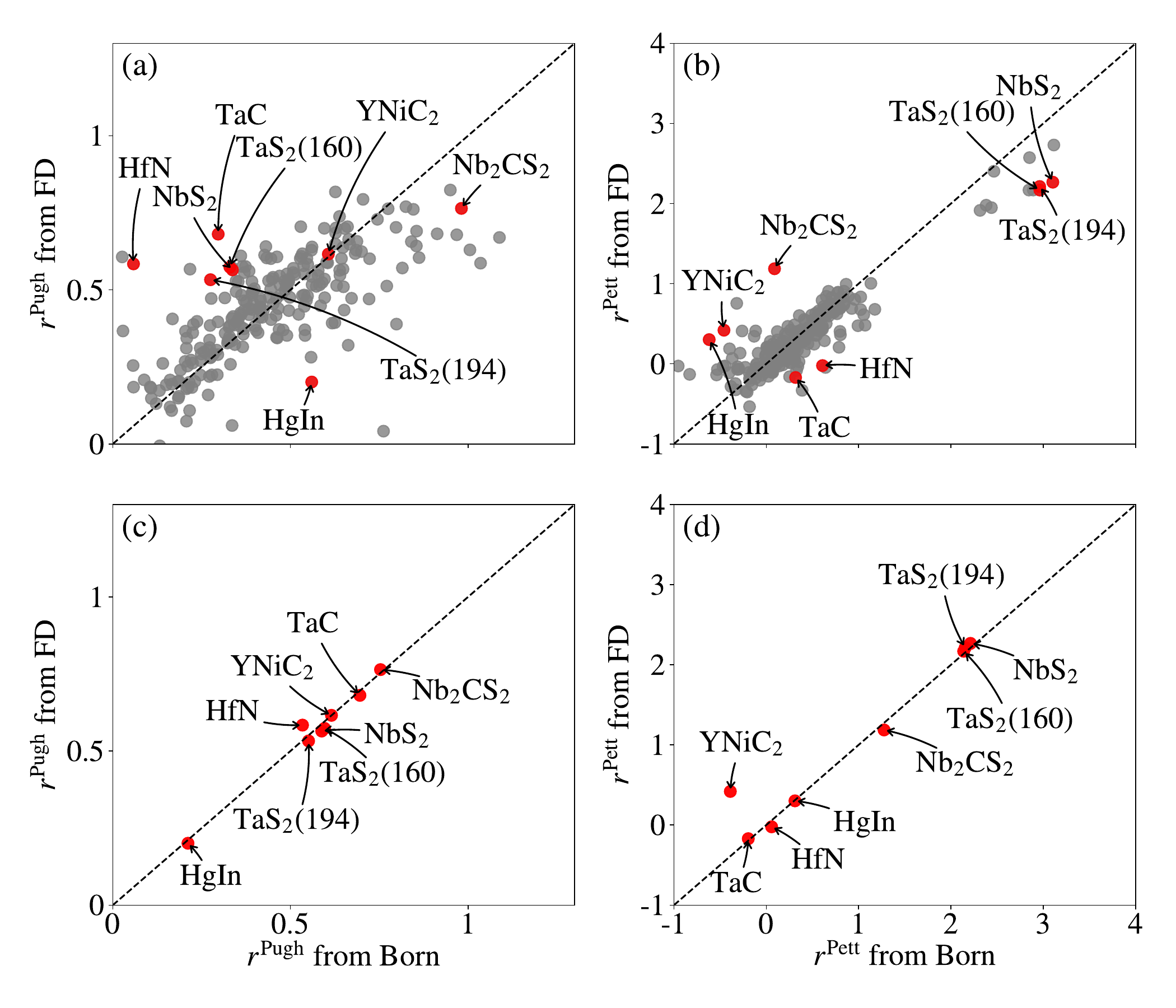}
    \caption{
    Comparison of (a) $r^{\rm Pugh}$=$\frac{G}{B}$ and (b) $r^{\rm Pett}$=$\frac{C_{12}-C_{44}}{B}$ for 250 predicted superconductors, obtained from the Born expansion and from finite-displacement (FD) calculations using the same computational parameters as Ref.~\cite{Bercx2025}.
    Panels (c) and (d) show the same quantities as (a, b) after densifying the $\mathbf{q}$-mesh and increasing the smearing from 20~mRy to 40~mRy for 8 selected outliers highlighted with red dots in (a,b).
    The dashed black lines indicate perfect agreement between FD and the Born expansion.
}
    \label{fig:fig1}
\end{figure}

To automate these calculations, we developed a new \textsc{AiiDA}~\cite{Huber2020,Uhrin2021} workflow called ``\href{https://github.com/aiidaplugins/aiida-mechanical}{\textsc{aiida-mechanical}}''~\cite{aiida-mechanical}. 
Figure~\ref{fig:fig1} compares the $r^{\rm Pugh}$ and  $r^{\rm Pett}$ obtained from the Born expansion and the FD method for the 250 materials.
We obtain Pearson correlation coefficients between these two methods of 0.471 and 0.900 for $r^{\rm Pugh}$ and $r^{\rm Pett}$, respectively.
This discrepancy comes from the $\mathbf{q}$-point sampling.
A convergence study across the database confirms that the Born expansion evaluated using a converged $\mathbf{q}$-point aligns with FD calculations, as illustrated in Sec.~\ref{sec:sections1}.
We selected eight materials with large discrepancies (highlighted as red dots in Fig.~\ref{fig:fig1}) which have correlations of 0.164 and 0.869, respectively. 
The evolution of their $r^{\rm Pugh}$, $r^{\rm Pett}$, and phonon band structures with respect to the $\mathbf{q}$-point sampling is shown in Sec.~\ref{sec:sections1}, revealing the convergence issue.

Using converged $\mathbf{q}$-point grids and increasing the smearing to 40~mRy, the agreement for these eight outliers improves.
As shown in Fig.~\ref{fig:fig1}, the Pearson correlations increase to 0.988 and 0.966 for $r^{\rm Pugh}$ and $r^{\rm Pett}$, respectively.
Because of the computational cost associated with the $\mathbf{q}$-point convergence, the FD method is more favorable for high-throughput computation.
Therefore, we proceed with the FD calculations for the remainder of this work.
We verify that the applied strains lie within the linear stress–strain regime for the FD calculations, as detailed in Sec.~\ref{sec:sections2}.

\begin{figure}[t]
	\centering
    \includegraphics[width=0.99\linewidth]{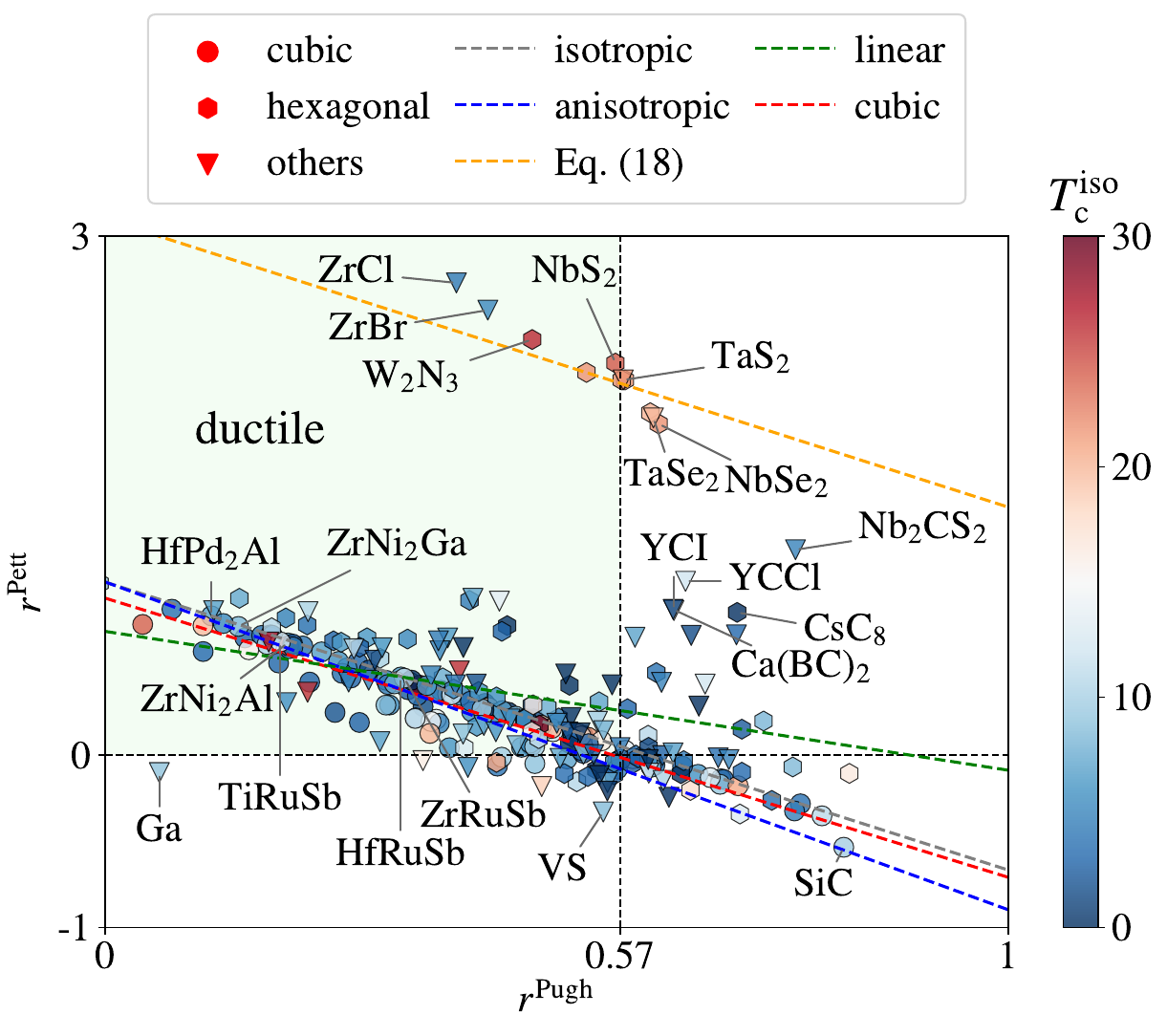}
	\caption{\label{fig:fig2}
    Correlation between $r^{\rm Pugh}$ and $r^{\rm Pett}$. 
    Ductile materials have $r^{\rm Pugh}<0.57$ and $r^{\rm Pett}>0$.
    The isotropic Migdal-Eliashberg superconductivity $T_{\rm c}^{\rm iso}$~(K) is shown with color and the crystal structure family with symbols.
    }
\end{figure}

In Fig.~\ref{fig:fig2}, we identify with a dashed green line a regression slope of -0.74 between the Pettifor ratio and Pugh's ratio.
This correlation can be analytically derived for cubic systems, where the two criteria are linked via the Zener anisotropy ratio $A \equiv 2C_{44}/(C_{11}-C_{12})$~\cite{Senkov2021},
\begin{equation}\label{eq:relation_pugh_pettifor}
	r^{\rm Pett} =  1 - \frac{5(2A+3)(3A+2)}{3(3A^2+19A+3)}\frac{G}{B},
\end{equation}
where for cubic materials with perfect isotropy and $A = 1$,
\begin{equation}\label{eq:relation_pugh_pettifor_iso}
	r^{\rm Pett} =  1 - \frac{5}{3} r^{\rm Pugh}.
\end{equation}

This relationship between Pugh's ratio and Pettifor's ratio is in close agreement with the linear relationship for the 113 cubic materials where the slope is -1.62, confirming the equivalence of these two criteria for highly symmetric structures~\cite{Senkov2021}.

\begin{table}[!ht]
\caption{
    Top 39 ductile superconductors with space group~(SG), bulk modulus (B), shear modulus (G), $r^{\rm Pugh}$ and $r^{\rm Pett}$.
    The superconductivity is reported with increasing level of theory from the Allen-Dynes transition temperature $T^{\rm AD}_{\rm c}$, to the isotropic Migdal-Eliashberg $T^{\rm iso}_{\rm c}$, to the anisotropic Migdal-Eliashberg $T^{\rm aniso}_{\rm c}$.
    The $c$ coefficient is the indicator for ductile superconductors, as defined in Eq.~\ref{eq:pugh_pettifor_tc_descriptor}.
    Superconducting values are from Ref.~\cite{Bercx2025}.
}
\small
\setlength{\tabcolsep}{1.9pt}
\begin{tabular}{lrrrrrrrrr}
\hline
Material   & SG & B & G & $r^{\rm Pugh}$ & $r^{\rm Pett}$ & $T^{\rm AD}_{\rm c}$ & $T^{\rm iso}_{\rm c}$ & $T^{\rm aniso}_{\rm c}$ & c\\
   &  & [GPa] & [GPa] &   &   & [K] & [K] & [K] & \\
\hline
W$_{2}$N$_{3}$ & 194 & 75.91 & 35.89 & 0.47 & 2.40 & 20.17 & 26.34 & 30.91 & 7.08 \\
NbS$_{2}$ & 194 & 25.02 & 14.13 & 0.56 & 2.27 & 18.52 & 24.31 & 25.30 & 6.36 \\
TaS$_{2}$ & 194 & 25.25 & 13.46 & 0.53 & 2.21 & 16.59 & 21.75 & 25.01 & 6.08 \\
ZrBr & 164 & 16.84 & 7.13 & 0.42 & 2.57 & 3.61 & 5.38 & 10.79 & 5.83 \\
TaSe$_{2}$ & 194 & 22.71 & 13.71 & 0.60 & 1.98 & 8.99 & 20.10 & 27.47 & 5.20 \\
NbSe$_{2}$ & 194 & 21.92 & 13.44 & 0.61 & 1.92 & 16.28 & 21.86 & 26.74 & 5.17 \\
Nb$_{3}$Sn & 223 & 162.42 & 32.94 & 0.20 & 0.63 & 18.30 & 25.61 & 35.55 & 3.28 \\
Nb & 229 & 168.45 & 18.23 & 0.11 & 0.75 & 15.11 & 19.66 & 21.18 & 3.27 \\
Nb$_{3}$Ge & 223 & 173.40 & 60.34 & 0.35 & 0.40 & 18.07 & 32.38 & 35.23 & 2.95 \\
PdH & 1 & 126.92 & 28.44 & 0.22 & 0.36 & 21.18 & 25.88 & 30.10 & 2.58 \\
NiS & 194 & 136.59 & 20.29 & 0.15 & 0.91 & 5.65 & 7.36 & 10.48 & 2.52 \\
RuO$_{2}$ & 136 & 255.47 & 100.18 & 0.39 & 0.49 & 21.85 & 26.05 & 34.02 & 2.51 \\
Ta$_{3}$Sn & 223 & 179.10 & 28.47 & 0.16 & 0.61 & 11.41 & 15.64 & 21.53 & 2.47 \\
V$_{4}$H & 141 & 183.11 & 41.12 & 0.22 & 0.83 & 7.97 & 10.38 & 11.48 & 2.41 \\
AgO & 15 & 103.03 & 12.36 & 0.12 & 0.83 & 5.43 & 6.86 & 7.56 & 2.37 \\
ZrS & 129 & 118.88 & 51.89 & 0.44 & 0.89 & 10.71 & 13.37 & 18.80 & 2.32 \\
RhSe & 194 & 139.09 & 25.64 & 0.18 & 0.74 & 8.26 & 10.79 & 12.21 & 2.31 \\
HfPd$_{2}$Al & 225 & 157.38 & 18.50 & 0.12 & 0.80 & 4.69 & 6.80 & 8.86 & 2.29 \\
TiRuSb & 216 & 142.43 & 27.63 & 0.19 & 0.65 & 9.98 & 12.88 & 13.57 & 2.26 \\
ZrNi$_{2}$Ga & 225 & 152.25 & 22.45 & 0.15 & 0.74 & 5.41 & 8.23 & 10.34 & 2.19 \\
TiNi & 221 & 158.47 & 28.38 & 0.18 & 0.64 & 8.06 & 10.35 & 11.11 & 2.04 \\
MoN & 187 & 346.64 & 166.70 & 0.48 & 0.17 & 26.70 & 32.06 & 40.36 & 2.04 \\
IrSi$_{2}$ & 225 & 181.55 & 23.73 & 0.13 & 0.76 & 4.14 & 5.32 & 5.61 & 2.03 \\
YCCl & 12 & 11.86 & 7.62 & 0.64 & 1.00 & 8.00 & 12.05 & 14.87 & 2.00 \\
IrTe$_{2}$ & 164 & 48.54 & 19.76 & 0.41 & 0.91 & 5.08 & 7.22 & 8.20 & 1.90 \\
Ni$_{3}$ZnN & 221 & 195.98 & 63.17 & 0.32 & 0.42 & 14.49 & 18.30 & 17.59 & 1.86 \\
YCI & 12 & 9.19 & 5.79 & 0.63 & 0.83 & 8.54 & 13.60 & 17.64 & 1.75 \\
NbCoSb & 216 & 155.70 & 59.03 & 0.38 & 0.35 & 16.22 & 20.41 & 21.40 & 1.73 \\
YPd$_{2}$Sn & 225 & 108.54 & 20.09 & 0.19 & 0.69 & 3.73 & 5.07 & 5.75 & 1.71 \\
ZrNi$_{2}$Al & 225 & 151.80 & 30.70 & 0.20 & 0.65 & 4.80 & 6.39 & 6.86 & 1.67 \\
RhTe & 194 & 119.28 & 31.16 & 0.26 & 0.65 & 4.92 & 6.37 & 7.57 & 1.54 \\
CoS & 194 & 162.39 & 49.43 & 0.30 & 0.56 & 7.97 & 10.17 & 14.71 & 1.54 \\
TiPtAl & 194 & 160.44 & 46.51 & 0.29 & 0.68 & 4.75 & 6.25 & 5.27 & 1.53 \\
Ti$_{3}$TlN & 221 & 141.47 & 33.88 & 0.24 & 0.60 & 5.23 & 6.75 & 5.77 & 1.49 \\
AgB$_{2}$ & 1 & 145.82 & 55.29 & 0.38 & 0.68 & 5.42 & 6.44 & 7.74 & 1.33 \\
In & 139 & 34.68 & 10.64 & 0.31 & 0.62 & 1.93 & 6.05 & 8.06 & 1.33 \\
TlBi$_{2}$ & 191 & 36.48 & 12.21 & 0.33 & 0.67 & 4.06 & 5.26 & 6.60 & 1.32 \\
Mo$_{3}$Os & 223 & 287.31 & 122.03 & 0.42 & 0.27 & 15.37 & 18.91 & 20.80 & 1.30 \\
ZrRuSb & 216 & 136.69 & 45.26 & 0.33 & 0.44 & 8.86 & 11.28 & 13.17 & 1.29 \\
\hline
\end{tabular}
\label{table:table1}
\end{table}

However, Fig.~\ref{fig:fig2} also reveals a severe breakdown of this linear correlation for a fraction of the dataset, primarily driven by lower-symmetry crystal structures such as hexagonal materials.
Therefore, we identify an approximate relationship for a group of trigonal and hexagonal materials, as indicated by the dashed orange line.
These compounds are predominantly layered materials with strong anisotropy:
\begin{equation}
\label{eq:layered_elastic_hierarchy}
C_{11},\, C_{12},\, C_{66}
\gg
C_{33},\, C_{13},\, C_{44}.
\end{equation}
The Voigt bulk and shear moduli as given by Eq.~\ref{eq:bulk_shear_modulus} can be approximated as:
\begin{align}
B^{\rm V}
&\approx
\frac{2(C_{11}+C_{12})}{9},
\\
G^{\rm V}
&\approx
\frac{7C_{11}-5C_{12}}{30}.
\end{align}
We observe that Reuss bulk and shear moduli are much smaller than their Voigt counterparts in these materials, the Hill averages can be approximated as:
\begin{align}
B
&=
\frac{B^{\rm V}+B^{\rm R}}{2}
\approx
\frac{B^{\rm V}}{2},
\\
G
&=
\frac{G^{\rm V}+G^{\rm R}}{2}
\approx
\frac{G^{\rm V}}{2}.
\end{align}
Consequently,
\begin{equation}
\label{eq:pugh_pettifor_layered_intercept}
r^{\rm Pett}
+
\frac{5}{3}r^{\rm Pugh}
\approx
\frac{7C_{11}+31C_{12}}
{4(C_{11}+C_{12})}.
\end{equation}
For this group of materials, the ratio $C_{12}/C_{11}$ is close to 0.3, which yields an intercept of 3.1:
\begin{equation}
\label{eq:pugh_pettifor_layered}
r^{\rm Pett}
\approx
3.1-\frac{5}{3}r^{\rm Pugh}.
\end{equation}
We also show in Fig.~\ref{fig:fig2} the value of isotropic Migdal-Eliashberg $T_{\rm c}^{\rm iso}$ with color but do not observe a clear trend between superconductivity and ductility.
Similar results are obtained for the Allen-Dyne $T_{\rm c}^{\rm AD}$ and anisotropic transition temperature $T_{\rm c}^{\rm aniso}$, as shown in Sec.~\ref{sec:sections3}.

Regarding ductility, materials with a Pugh's ratio below 0.57 and a positive  Pettifor's ratio are considered ductile, which corresponds to the top left quadrant of Fig.~\ref{fig:fig2}. 
To select promising materials which balance a low Pugh's ratio, a high Pettifor ratio, and a high superconducting transition temperature, we propose the following indicator $c$ for superconductor ductility:
\begin{equation}\label{eq:pugh_pettifor_tc_descriptor}
    c \equiv \frac{T_{\rm c}^{\rm iso}}{T_{\rm c}^{\rm iso, avg}} + \frac{ r^{\rm Pett}}{r^{\rm Pett, avg}}-\frac{r^{\rm Pugh} }{r^{\rm Pugh, avg}},
\end{equation}
where the denominators are quantities averaged over the 250 metals.
A summary of the top 39 superconductors sorted by their superconductor ductility $c$ is presented in Table~\ref{table:table1}.

The promising superconductors encompass a broad spectrum of structure types represented in the source \textsc{supercond-EPW} database. 
Specifically, the database contains 68 hexagonal, 113 cubic and 69 other structures.
We find that a large number of the highly ductile candidates are hexagonal.
However, Pugh's ratio and Pettifor's ratio were originally formulated and predominantly validated for cubic and relatively isotropic materials. 
Their predictive reliability for anisotropic structures, such as hexagonal systems, is limited. 
Furthermore, evaluating the microscopic plasticity of hexagonal materials is notoriously complex. 
Unlike cubic systems, which possess well-defined and highly symmetric primary slip systems, hexagonal materials exhibit competing basal, prismatic, and pyramidal slip systems that depend sensitively on the $c/a$ lattice parameter ratio. 
A comprehensive evaluation of all these potential slip systems is computationally prohibitive and beyond the scope of the current systematic screening.
Therefore, for the subsequent analysis of the stacking fault energy, we restrict our focus to cubic systems. 
This selection is also inspired by our previous superconductivity study~\cite{Bercx2025}, where we found exceptional half-Heusler superconductors with a valence electron count~(VEC) of 17.
\begin{table}[!t]
\caption{
    Relaxed generalized stacking fault and surface energies for simple face-centered cubic metals on the (111) slip plane along the [11$\bar{2}$] direction.
    $\gamma^{\rm USFE}$ denotes the unstable stacking fault energy;
    $\gamma^{\rm ISFE}$ and $\gamma^{\rm ESFE}$ represent the intrinsic and extrinsic stacking fault energies, respectively; 
    $\gamma^{\rm UTE}$ is the unstable twinning energy; 
    and $\gamma^{\rm surface}$ indicates the surface energy.
    All energy values are reported in (mJ/m$^2$).
    }
\small
\setlength{\tabcolsep}{3.5pt}
\begin{tabularx}{\columnwidth}{l l l l l l}
\toprule
 & $\gamma^{\rm USFE}$ & $\gamma^{\rm ISFE}$ & $\gamma^{\rm UTE}$ & $\gamma^{\rm ESFE}$ & $\gamma^{\rm surface}$ \\
\midrule
Ag           & 91.89                            & 14.45                            & 99.78                        & 16.65                            & 722.06                    \\
             & 97~\cite{Tu2019}                 & 21$\pm$7~\cite{Dillamore1965}    & 105~\cite{Bernstein2004}     & 29~\cite{Hartford1998}           & 770~\cite{Tu2019}         \\
             & 190$\pm$15~\cite{Bernstein2004}  & 29~\cite{Hartford1998}           &                              & 33$\pm$5~\cite{Bernstein2004}    &                           \\
             &                                  & 35$\pm$15~\cite{Bernstein2004}   &                              &                                  &                           \\
\addlinespace
Al           & 166.47                           & 117.84                           & 210.73                       & 133.68                           & 844.12                    \\
             & 83-190~\cite{Zimmerman2000}      & 280$\pm$50~\cite{Dillamore1965}  & 207~\cite{Bernstein2004}     & 147~\cite{Hartford1998}          & 800~\cite{Tu2019}         \\
             & 199$\pm$25~\cite{Bernstein2004}  & 6-170~\cite{Zimmerman2000}       &                              & 184$\pm$76~\cite{Bernstein2004}  &                           \\
             &                                  & 161~\cite{Hartford1998}          &                              &                                  &                           \\
             &                                  & 146~\cite{Tu2019}                &                              &                                  &                           \\
             &                                  & 203$\pm$77~\cite{Bernstein2004}  &                              &                                  &                           \\
\addlinespace
Au           & 64.96                            & 21.10                            & 75.39                        & 22.95                            & 714.12                    \\
             & 56~\cite{Tu2019}                 & 52$\pm$15~\cite{Dillamore1965}   & 135~\cite{Bernstein2004}     & 44~\cite{Bernstein2004}          & 740~\cite{Tu2019}         \\
             &                                  & 44~\cite{Bernstein2004}          &                              &                                  &                           \\
\addlinespace
Cu           & 222.35                           & 3.46                             & 224.83                       & 11.01                            & 1203.99                   \\
             & 158-210~\cite{Zimmerman2000}     & 85$\pm$30~\cite{Dillamore1965}   & 236~\cite{Bernstein2004}     & 54~\cite{Hartford1998}           & 1310~\cite{Tu2019}        \\
             & 160~\cite{Tu2019}                & 27-49~\cite{Zimmerman2000}       &                              & 63$\pm$10~\cite{Bernstein2004}   &                           \\
             & 184$\pm$26~\cite{Bernstein2004}  & 53~\cite{Hartford1998}           &                              &                                  &                           \\
             &                                  & 54$\pm$16~\cite{Bernstein2004}   &                              &                                  &                           \\
\addlinespace
Pb           & 75.42                            & 44.56                            & 96.27                        & 37.66                            & 361.38                    \\
             & 98~\cite{Bernstein2004}          & 30~\cite{Bernstein2004}          & 108~\cite{Bernstein2004}     & 31~\cite{Bernstein2004}          &                           \\
\addlinespace
Pt           & 285.50                           & 285.95                           & 378.91                       & 284.66                           & 1504.09                   \\
             & 388~\cite{Bernstein2004}         & 322~\cite{Bernstein2004}         & 521~\cite{Bernstein2004}     & 316~\cite{Bernstein2004}         &                           \\
\bottomrule
\end{tabularx}
\label{table:table2}
\end{table}
Pugh's and Pettifor's ratios rely on the elastic properties of materials. 
However, the elastic parameters cannot capture the localized nature of plastic deformation, often resulting in overestimated theoretical shear stresses. 
Fundamentally, a material's intrinsic ductility is governed by the competition between crack propagation and dislocation emission.
Rice's ratio evaluates the energetic penalty of nucleating a dislocation compared to creating a new free fracture surface.
To accurately evaluate this competition, we use Rice's criterion~\cite{Rice1974}, which provides an indicator of ductility beyond the linear regime.
Specifically, the maximum of the GSFE along the selected slip direction represents the energy barrier that must be overcome to rigidly shift two crystal halves and initiate dislocation emission.
The characteristics of the GSFE are highly dependent on crystal structures and slip planes.
Distinct structures possess unique slip systems, leading to specific stacking sequences and, consequently, various types of stacking faults.

To automate the calculation of stacking fault energies, we use the \href{https://github.com/aiidaplugins/aiida-mechanical}{\textsc{aiida-mechanical}} package, which currently supports the following specific crystal classes: $A_1$, $A_2$, $B_1$, $B_2$, $C1_{b}$, and $\rm{L2}_1$.
In this workflow, we first construct a supercell perpendicular to the slip plane.
We then do constrained out-of-plane relaxation.
Detailed descriptions of the tilted-cell algorithm, stacking sequences, the specific slip systems for each crystal class, and the relaxation method are provided in Sec.~\ref{sec:sections4}.

The unstable stacking fault energies are calculated as:
\begin{equation}\label{eq:usfe}
    \gamma^{\rm USFE} \equiv \frac{E^{\rm USF}-E^0}{A},
\end{equation}
where $E^{\rm USF}$ and $E^0$ represent the total energies of the supercell with and without the unstable stacking fault, respectively, and $A$ denotes the area of the fault plane. 
The surface energies $\gamma^{\rm surface}$ are calculated as:
\begin{equation}\label{eq:surface_energy}
    \gamma^{\rm surface} \equiv \frac{E^{\rm slab}-E^0}{2A},
\end{equation}
where $E^{\rm slab}$ is the total energy of the cleaved vacuum slab.
Both the energy and the area depend on the specific slip plane chosen.
Rice's ratio is then defined as:
\begin{equation}\label{eq:rice}
    r^{\rm Rice} \equiv \frac{\gamma^{\rm USFE}}{\gamma^{\rm surface}} = \frac{2(E^{\rm USF}-E^0)}{E^{\rm slab}-E^0}.
\end{equation}

When such information beyond the linear regime is known, we propose an improved indicator for ductile superconductor $c^*$ that includes the original indicator $c$ and $r^{\rm Rice}$:
\begin{equation}\label{eq:pugh_pettifor_tc_rice_descriptor}
    c^* \equiv c - \frac{ r^{\rm Rice}}{r^{\rm Rice, avg}}.
\end{equation}

We first benchmark our workflow on several elemental metals with face-centered cubic~(FCC, A$_1$) structures.
In FCC systems, the relevant slip for stacking fault formation occurs on the \{111\} plane along the $[11\bar{2}]$ direction with a Burgers vector of $\textbf{b}=\frac{a}{6}[11\bar{2}]$. 
The perfect crystal exhibits a stacking sequence of ABCABCABC. 
A shear displacement of the upper half of the crystal by $\textbf{b}$ transforms the sequence into ABCBCABCA, resulting in an intrinsic stacking fault~(ISF), with its formation energy defined as:
\begin{equation}\label{eq:isfe}
    \gamma^{\rm ISFE} \equiv \frac{E^{\rm ISF}-E^0}{A}.
\end{equation}
The energy barrier encountered during this transition is the unstable stacking fault energy, which, alongside the surface energy, is critical for evaluating Rice's criterion.
Starting from the ISF structure, a further displacement of the upper layers while keeping one additional layer fixed produces the sequence ABCBABCAB.
The corresponding formation energy is the extrinsic stacking fault energy~(ESFE):
\begin{equation}\label{eq:esfe}
    \gamma^{\rm ESFE} \equiv \frac{E^{\rm ESF}-E^0}{A}.
\end{equation}
The energy barrier encountered between the ISF and ESF states is termed the unstable twinning energy~(UTE):
\begin{equation}\label{eq:ute}
    \gamma^{\rm UTE} \equiv \frac{E^{\rm UT}-E^0}{A}.
\end{equation}

\begin{table}[htbp]
\caption{
    Relaxed generalized stacking fault and surface energies for typical simple body-centered cubic metals (Li, Na, Nb, V) and rocksalt materials (KCl, NaCl) along the (1$\bar{1}0$) slip plane.
    $\gamma^{\rm USFE}$ is the unstable stacking fault energy
    and $\gamma^{\rm surface}$ is the surface energy.
    }
\small

\newcolumntype{L}{>{\hsize=1.5\hsize\raggedleft\arraybackslash}X} %
\newcolumntype{S}{>{\hsize=0.81\hsize\raggedleft\arraybackslash}X} %
\begin{tabularx}{\columnwidth}{l r@{\hspace{3.7em}} r@{\hspace{3.7em}} r}
\toprule
 Material & \multicolumn{2}{c}{$\gamma^{\rm USFE}$ (mJ/m$^2$)} & $\gamma^{\rm surface}$  (mJ/m$^2$) \\
\cmidrule(lr){2-3}
 & This work & Reference & \\
 \midrule
Li & 77.86 & 71 \cite{Zhang2023} & 495.72 \\
Na & 49.96 & 48 \cite{Zhang2023} & 208.86 \\
Nb & 698.61 & 674 \cite{Zhang2023} & 2117.88 \\
V & 715.01 & 713 \cite{Zhang2023} & 2494.39 \\
KCl & 154.59 & 193.8 \cite{Liu2012} & 267.47 \\
NaCl & 177.34 & 225.2 \cite{Liu2012} & 350.96 \\
\bottomrule
\end{tabularx}
\label{table:table3}
\end{table}
The calculated relaxed ISFE, ESFE, USFE, and UTE for representative FCC metals (Al, Cu, Ag, Pt, Au, and Pb) are summarized and compared with prior studies in Table~\ref{table:table2}.
The overall agreement is good and falls within the variability reported in previous works.
We also provide the unrelaxed (frozen) stacking fault results in Sec.~\ref{sec:sections4}.
The GSFEs on the (111) plane for A$_1$ metals are highly sensitive to the $\textbf{k}$-point sampling and convergence threshold of the total energy.
Using an insufficient $\textbf{k}$-point sampling or low convergence threshold can lead to unphysical results, such as negative ISFE (as observed in Ag in Sec.~\ref{sec:sections4}) or inconsistent total energies between the pristine structure and the structure shifted by a full unit vector.
The detailed convergence study for the stacking fault energies is provided in Sec.~\ref{sec:sections4}.

In BCC systems, slip typically occurs on the (1$\bar{1}$0) planes along the [111] direction with a Burgers vector of $\textbf{b} = \frac{a}{2}\langle 111 \rangle$.
The unstable stacking fault energy corresponds to the energy barrier encountered at a displacement of $\frac{1}{2}\textbf{b}$.
The calculated $\gamma^{\rm USFE}$ values and corresponding references for Li, Na, V, and Nb are listed in Table~\ref{table:table3}.
For the ionic B$_1$ crystals~(NaCl, KCl), the easiest slip plane is (1$\bar{1}$0)~\cite{Liu2012}. 
We note that the spacing between adjacent atomic layers normal to this plane is $\sqrt{2}a/4$, which is smaller than that of the (001) plane ($a/2$). 
To obtain a reasonable USFE, structural relaxation along the [1$\bar{1}$0] direction is necessary to release the excessive stress at the faulted plane. 
Omitting this structural relaxation leads to an order-of-magnitude overestimation of the USFE for ionic B$_1$ crystals.
Details regarding these convergence tests and the critical role of structural relaxation for both BCC and B$_1$ systems are provided in Sec.~\ref{sec:sections4}.
These benchmarks confirm that the current workflow provides a framework for evaluating the ductility of new superconducting candidates.

Expanding our study to the promising superconductors listed in Table~\ref{table:table1}, we investigated the following systems:
(i) $\rm{C1}_b$ half-Heusler alloys, including \ce{NbCoSb}, \ce{TiRuSb}, and \ce{ZrRuSb};
and (ii) $\rm{L2}_1$ full-Heusler alloys, including \ce{HfPd2Al}, \ce{ZrNi2Al}, \ce{YPd2Sn}, and \ce{ZrNi2Ga}, as shown in Table~\ref{table:table4}.
Although the elastic descriptors $r^{\rm Pugh}$ and $r^{\rm Pett}$ are strongly correlated with each other, their correlation with the Rice ratio is weak for this seven-material set.
Among the half-Heusler and full-Heusler systems, the (1$\bar{1}$0) plane is the easiest slip plane compared to (100) and (111).
For the three common slip directions [110], [001], and [112], we observe $\gamma^{\rm USFE, [110]} > \gamma^{\rm USFE, [112]} > \gamma^{\rm USFE, [001]}$ for the relaxed configurations.
In addition, antiphase boundaries are observed along the [112] and [001] directions.

The out-of-plane relaxation reduces the unstable stacking fault energies by approximately 50\%.

\begin{table}[!b]
\caption{
    Relaxed unstable stacking fault and surface energies for half-Heusler and full-Heusler ductile superconductor candidates along the (1$\bar{1}$0) slip plane.
    The $r^{\rm Rice}$, $r^{\rm Pugh}$, $r^{\rm Pett}$, and superconductor ductility $c^*$ are reported.
    All energy values are reported in (mJ/m$^2$).
    }

\setlength{\tabcolsep}{4.5pt} 

\begin{tabularx}{\columnwidth}{l S[table-format=3.3] S[table-format=4.3] *{4}{S[table-format=1.3]}}
\toprule
 Material & $\gamma^{\rm USFE}$ & $\gamma^{\rm surface}$ & $r^{\rm Rice}$ & $r^{\rm Pugh}$ & $r^{\rm Pett}$ & $c^*$\\
\midrule
\ce{HfPd2Al} & 674.26 & 1545.05 & 0.44  & 0.12 & 0.80  & 1.41 \\
\ce{TiRuSb}  & 701.26 & 1495.03 & 0.47  & 0.19 & 0.65  & 1.32 \\
\ce{ZrNi2Ga} & 910.81 & 1651.06 & 0.55  & 0.15 & 0.74  & 1.09 \\
\ce{NbCoSb}  & 617.47 & 1343.65 & 0.46  & 0.38 & 0.35  & 0.81 \\
\ce{ZrNi2Al} & 887.42 & 1779.73 & 0.50  & 0.20 & 0.65  & 0.67 \\
\ce{YPd2Sn}  & 630.80 & 1057.52 & 0.60  & 0.19 & 0.69  & 0.51 \\
\ce{ZrRuSb}  & 686.99 & 1431.02 & 0.48  & 0.33 & 0.44  & 0.33 \\
\bottomrule
\end{tabularx}
\label{table:table4}
\end{table}

A detailed comparison between the unrelaxed and relaxed results, alongside the convergence studies for the stacking fault energies, is provided in Sec.~\ref{sec:sections4}.
Additionally, to ensure the accuracy of $r^{\rm Rice}$, the surface energies for these seven materials are converged with respect to the vacuum spacing, as detailed in Sec.~\ref{sec:sections5}.
The periodic $\gamma^{\rm GSFE}$ is fitted using the following Fourier series:
\begin{align}
    \gamma^{\text{GSFE,periodic}}(b) &= 
    \sum_{i=1}^{4} 
    \left \{ A_i \sin(2i\pi b) + B_i \left[ \cos(4i\pi b) - 1 \right] \right \},
\end{align}
where $A_i$ and $B_i$ are the four fitting coefficients.
 
Among the studied materials, \ce{ZrRuSb} is of particular interest.
As noted in Ref.~\cite{Bercx2025}, it is an exceptional 17-VEC half-Heusler superconductor. We here find that it also exhibits a relatively low USFE of 686.99~mJ/m$^2$ and a Rice's ratio of 0.48.
However, \ce{TiRuSb} possesses the highest predicted superconductor ductility indicator ($c^* = 1.32$) among the studied half-Heuslers. 
Although its USFE (701.26~mJ/m$^2$) is slightly higher than that of \ce{ZrRuSb}, its proportionally higher surface energy yields a lower Rice's ratio of 0.47.
We thus conclude that \ce{TiRuSb} is also a promising half-Heusler ductile superconductor candidate.
\begin{figure}[t]
	\centering
    \includegraphics[width=0.85\linewidth]{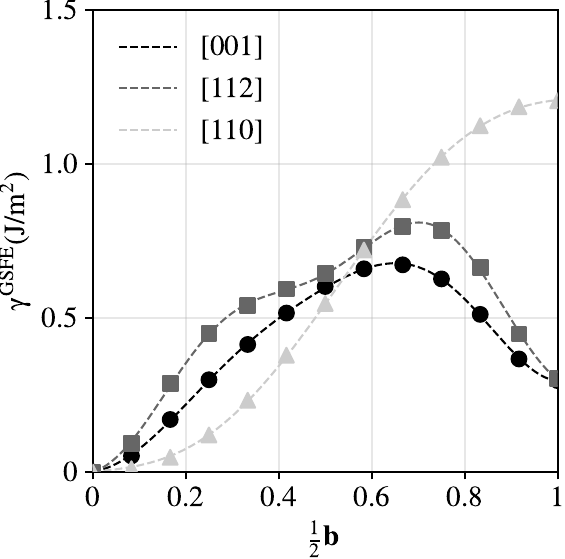}
	\caption{\label{fig:fig3}
    Generalized stacking fault energy $\gamma^{\rm GSFE}$ for the L2$_1$-type full-Heusler material \ce{HfPd2Al} on the (1$\bar{1}$0) plane. 
    The slip directions [110], [112], [001] are compared.
    }
\end{figure}

For full-Heusler materials, the easiest slip plane remains (1$\bar{1}$0) but the barriers along the three directions follow $\gamma^{[110], \rm USFE} > \gamma^{[112], \rm USFE} > \gamma^{[001], \rm USFE}$.
Based on this, we find that \ce{HfPd2Al} and \ce{ZrNi2Ga} are  promising full-Heusler ductile superconductors.
\ce{HfPd2Al} possesses the most favorable $r^{\rm Rice} = 0.44$.
This result is consistent with its favorable elastic descriptors.
It has the lowest Pugh ratio $r^{\rm Pugh}  = 0.12$ and the highest Pettifor ratio $r^{\rm Pett} = 0.80$ among the full-Heusler candidates.
Conversely, \ce{ZrNi2Ga} exhibits a slight discrepancy between microscopic and macroscopic indicators.
While its elastic properties suggest high ductility ($r^{\rm Pugh} = 0.15$, $r^{\rm Pett} = 0.74$), its higher Rice's ratio (0.55) indicates that dislocation emission is less favorable relative to cleavage.
This highlights the critical importance of GSFE calculations to verify ductility and demonstrates that GSFE calculations are necessary as an additional screening  beyond elastic descriptors.

\section{Conclusion}
In this work, we studied the mechanical properties of 250 potential superconductors identified through high-throughput electron-phonon coupling calculations.
By combining finite-displacement elastic constant calculations with rigorous generalized stacking fault energy calculations, we established a multi-scale screening workflow that bridges the gap between superconducting transition temperatures and ductility for engineering applications.
Our elastic constant analysis revealed that while the Born expansion offers a low-cost alternative, the finite-displacement method remains more robust for diverse crystal structures without requiring extensive $\mathbf{q}$-point convergence tests.
Based on Pugh's and Pettifor's criteria, we proposed a subset of candidates exhibiting both high predicted \Tc~and elastic compliance.
Further microscopic refinement using GSFE calculations and Rice's ratio highlights \ce{HfPd2Al}, \ce{TiRuSb}, and \ce{ZrNi2Ga}  as the most promising ductile superconductors.
Notably, the 17-VEC superconductor \ce{ZrRuSb} also demonstrates low energy barriers for plastic deformation and a balanced Rice's ratio.
It exhibits a more favorable balance between slip and cleavage than typical brittle intermetallics.

\section*{Acknowledgments}
We thank Dr. Changpeng Lin for providing an early version of the \textsc{Quantum ESPRESSO} code for the elastic property calculations~\cite{Lin2022,Lin2026}.
We also thank Prof.~Nicola Marzari and Dr.~Marnik Bercx for useful discussions.
S. P. is a Research Associate of the Fonds de la Recherche Scientifique - FNRS.
This work was supported by the Fonds de la Recherche Scientifique - FNRS under Grants number T.0183.23 (PDR) and  T.W011.23 (PDR-WEAVE). 
This publication was supported by the Walloon Region in the strategic axe FRFS-WEL-T.
Computational resources have been provided by the EuroHPC JU award granting access to MareNostrum5 at Barcelona Supercomputing Center (BSC), Spain (Project ID: EHPC-EXT-2023E02-050), by the Consortium des Équipements de Calcul Intensif (CÉCI), funded by the FRS-FNRS under Grant No. 2.5020.11 and by the Walloon Region, by the Tier-1 supercomputer of the Walloon Region (Lucia) with infrastructure funded by the Walloon Region under the grant agreement n°1910247, and by the Belgian share of the EuroHPC LUMI supercomputer~(Project ID: UCLouvain-MODL-perovskite-T0477).

\clearpage

\twocolumn[
\begin{@twocolumnfalse}

\begin{center}
    {\Large Supplementary Information: In search of novel ductile superconductors}
    \vspace{1.5em}

    \vspace{1.5em}

    {\normalsize
    Yiming Zhang$^{a}$, Samuel Ponc\'e$^{a,b,*}$
    \par}

    \vspace{0.8em}

    {\scriptsize \it
    $^a$European Theoretical Spectroscopy Facility, Institute of Condensed Matter and Nanosciences, Université catholique de Louvain, Chemin des Étoiles 8, B-1348 Louvain-la-Neuve, Belgium. 	
    \par
    $^b$WEL Research Institute, avenue Pasteur 6, 1300 Wavre, Belgium.		
    }

    \vspace{0.5em}

\end{center}
\end{@twocolumnfalse}
]

\setcounter{section}{0}
\setcounter{figure}{0}
\setcounter{table}{0}
\setcounter{equation}{0}

\renewcommand{\thesection}{S\arabic{section}}
\renewcommand{\thefigure}{S\arabic{figure}}
\renewcommand{\thetable}{S\arabic{table}}
\renewcommand{\theequation}{S\arabic{equation}}

\section{Convergence of the \texorpdfstring{$q$}{q}-point distance}
\label{sec:sections1}

\begin{figure}[!htb]
    \centering
    \includegraphics[width=\linewidth]{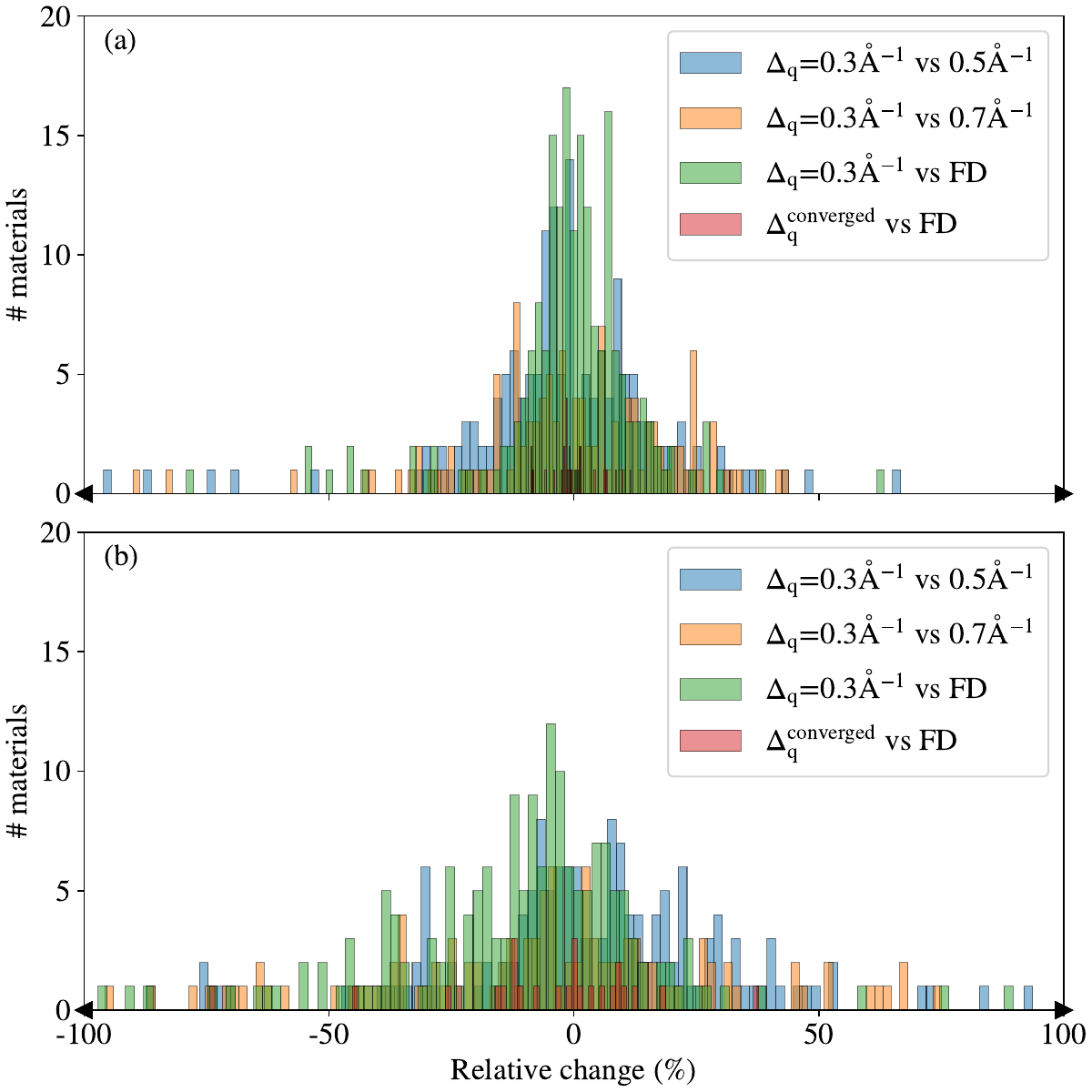}
    \caption{
        Comparison of the (a) bulk modulus $B$ and (b) shear modulus $G$ calculated at $\mathbf{q}$-point distance of \absq=$0.3$, $0.5$, and $0.7~{\AA}^{-1}$.
        Both the complete dataset at \absq=$0.3~{\AA}^{-1}$ and the converged subset are compared against the finite-difference (FD) calculations.
    }
    \label{fig:figs1}
\end{figure}

In Fig.~\ref{fig:figs1}, we investigate the convergence behavior of the bulk modulus $B$ and shear modulus $G$ with respect to the $\mathbf{q}$-point distances used in the density functional perturbation theory (DFPT) calculations.
The \textsc{Quantum ESPRESSO}~\cite{Giannozzi2017} screening underlying the \textsc{supercond-EPW}~\cite{Bercx2025} database contains 4093, 1380, and 578 phonon workchains computed at $\mathbf{q}$-point distances of \absq=$0.7$, $0.5$, and $0.3~{\AA}^{-1}$, respectively.
This dataset enables a statistical assessment of the robustness of the Born expansion approach across a diverse range of materials.
For the 250 predicted superconductors, we compare their DFPT results at \absq=$0.5~{\AA}^{-1}$ with FD calculations. 
The latter are performed using \textsc{thermo\_pw}~\cite{thermopw} by applying strains of $\pm 0.0025$ and $\pm 0.0075$.
Both computational sets share identical density functional theory (DFT) parameters (pseudopotentials, cutoffs, and $\mathbf{k}$-meshes), ensuring a strict benchmark.

\vspace{0.5\baselineskip}

To minimize the influence of unconverged phonons on the elastic properties, we analyse the relative changes in the moduli across the three $\mathbf{q}$-point distances. 
We find that 239 bulk moduli and 158 shear moduli converge within a relative change threshold of 5\%.
Among these converged materials, 73 bulk and 57 shear moduli belong to the 250 superconductors. 
Furthermore, out of the 578 structures computed with the finest DFPT $\mathbf{q}$-point grid (\absq=$0.3~{\AA}^{-1}$), 197 materials were identified as dynamically stable. 
Without strict $\mathbf{q}$-point convergence, the DFPT-predicted moduli for metals often fail to reach the high precision established in previous benchmarks (such as the study by Lin \textit{et al.}~\cite{Lin2026} on silicon). 
When we apply the 5\% $\mathbf{q}$-point convergence criterion to these dynamically stable materials, only 30 compounds remain in agreement. 
This significant reduction indicates that the Born expansion approximation is highly sensitive to the chosen $\mathbf{q}$-point density, particularly for metallic systems.

\begin{figure}[H]
    \centering
    \includegraphics[width=0.99\columnwidth]{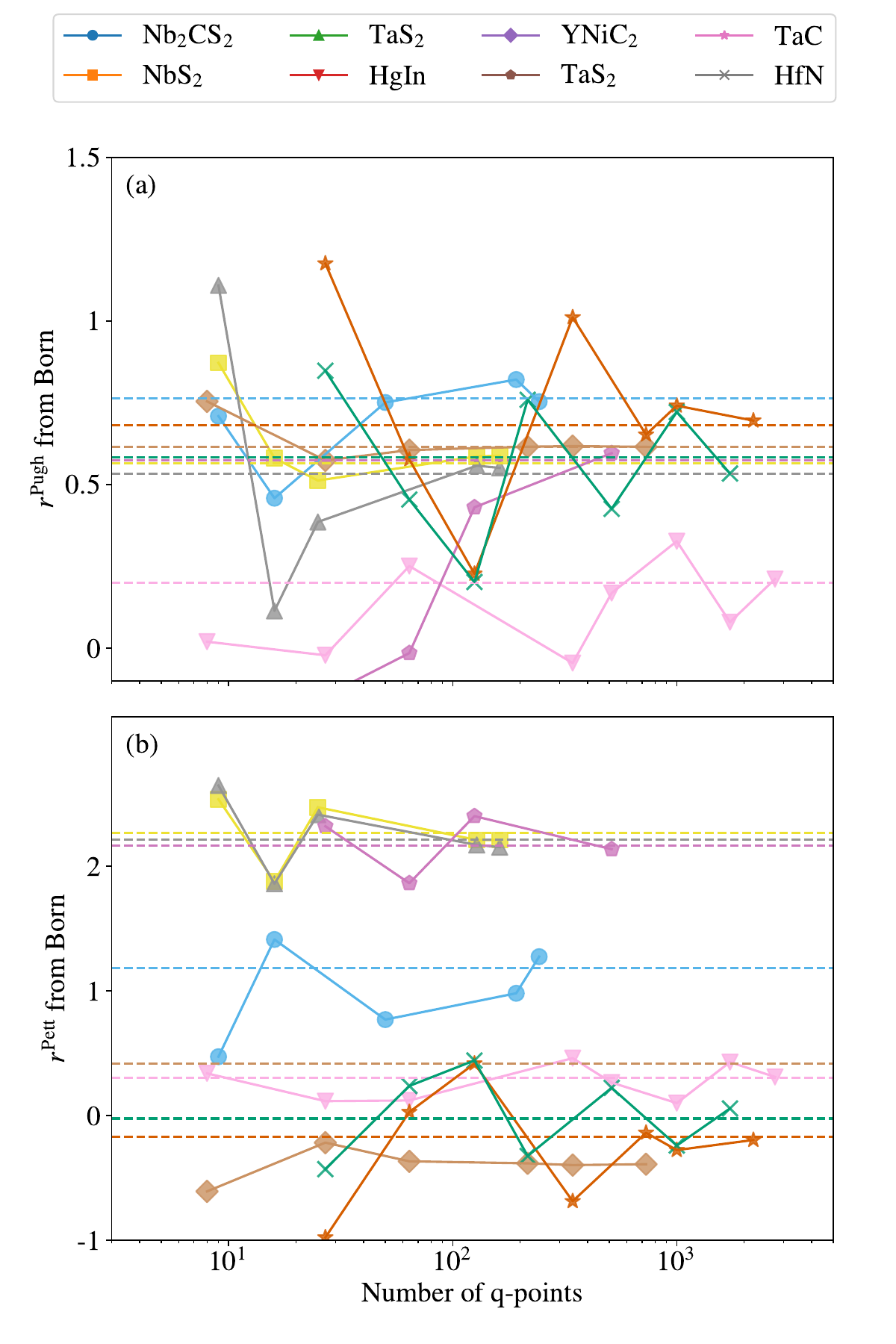}
    \caption{
        Convergence of (a)~$r^{\rm Pugh}$ and (b)~$r^{\rm Pett}$ for eight selected materials computed using progressively finer $\mathbf{q}$-point distances.
        The x-axis reports the number of $\mathbf{q}$-points in the full Brillouin zone and the densest grid is 16$\times$16$\times$16.
        The horizontal dashed lines indicate reference values from FD calculations using \textsc{thermo\_pw}.
        }
    \label{fig:figs2}
\end{figure}

\begin{figure}[H]
    \centering
    \includegraphics[width=0.9\columnwidth]{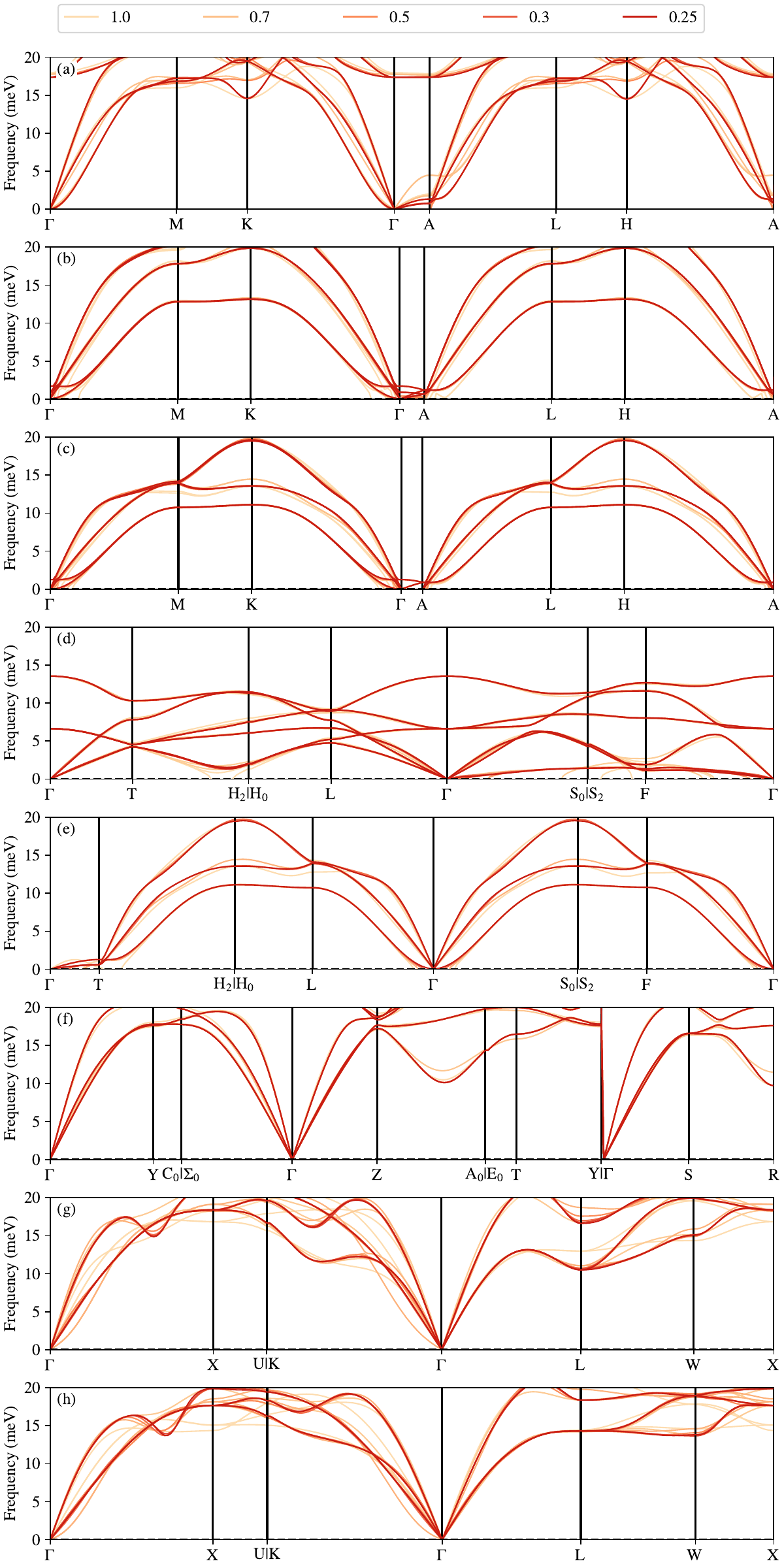}
    \caption{
       Convergence of the phonon band structures with respect to the $\mathbf{q}$-point distance~(${\AA}^{-1}$) for (a)~\ce{Nb2CS2}, (b)~\ce{NbS2}, (c)~\ce{TaS2}~(space group 194), (d)~\ce{HgIn}, (e)~\ce{YNiC2}, (f)~\ce{TaS2}~(space group 164), (g)~\ce{TaC}, and (h)~\ce{HfN}.
       The densest $\mathbf{q}$-point grid is 16$\times$16$\times$16 and we used a 40~mRy smearing.
        }
    \label{fig:figs3}
\end{figure}

To investigate these discrepancies between the Born expansion approach and the FD method, we selected eight representative outliers for a manual $\mathbf{q}$-point convergence study: \ce{HfN}, \ce{TaC}, \ce{TaS2} (space group 194), \ce{TaS2} (space group 164), \ce{NbS2}, \ce{YNiC2}, \ce{Nb2CS2}, and \ce{HgIn}.
As shown in Fig.~\ref{fig:figs2}, we track $r^{\rm Pugh}$ and $r^{\rm Pett}$ using progressively finer $\mathbf{q}$-point grids and compare them with the FD reference calculations (marked by horizontal dashed lines).
The evolution of the phonon band structures for these eight materials is presented in Fig.~\ref{fig:figs3}. 
The poor convergence of the elastic properties in these candidates can be attributed to Kohn anomalies~\cite{Enrico2025}, which induce sharp dips in the phonon dispersions.
To verify this hypothesis, we performed calculations using a larger smearing of 40~mRy to smooth out the Fermi surface nesting. 
As discussed in the main manuscript, suppressing these Kohn anomalies improves both the convergence and the alignment between the DFPT and FD results, confirming the sensitivity of the Born expansion approach to the $\mathbf{q}$-point grids.

\FloatBarrier
\section{Convergence of the finite-displacement calculations}
\label{sec:sections2}

In FD calculations, the elastic tensor is obtained by linearly fitting the stress-strain response for a set of strains of $\pm 0.0025$ and $\pm 0.0075$.
Therefore, it is essential to verify that these applied strains keep the system strictly within the linear elastic regime.

\begin{figure}[!htb]
    \centering
    \includegraphics[width=0.99\columnwidth]{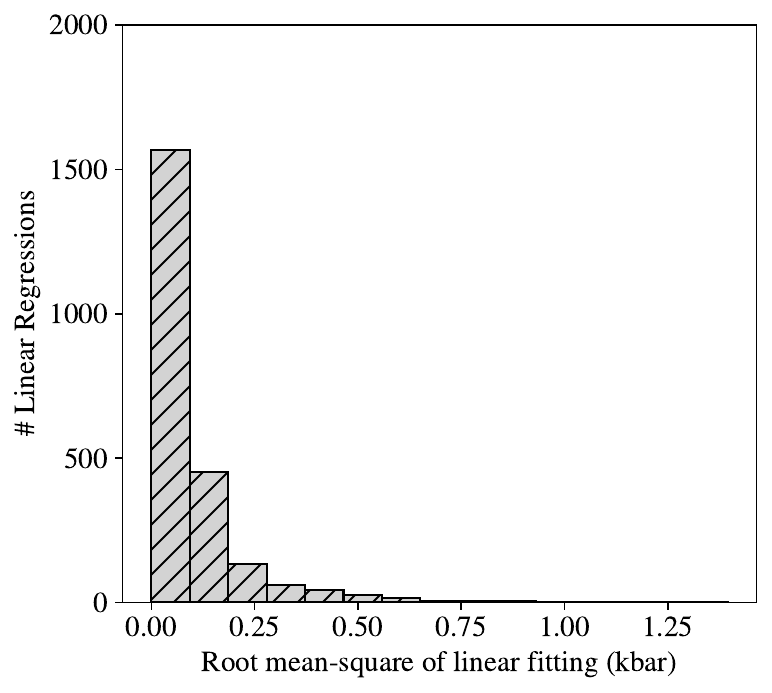}
    \caption{Convergence tests for the finite-displacement (FD) elastic workflow showing the distribution of root mean square~(RMS) errors for each independent elastic tensor component across the 298 materials~(2319 fits in total).}
    \label{fig:figs4}
\end{figure}

To assess the quality of these linear fits, we extract the root mean-square (RMS) error for every linear regression performed on all of the independent elastic tensor entries across the screened candidates.
Figure~\ref{fig:figs4} illustrates the statistical distribution of these RMS errors.
The histogram shows that the vast majority of the linear regressions have small RMS errors, below 0.125~kbar.
This confirms that the applied strains successfully maintain the materials within their linear limits.
Consequently, the derived elastic constants and the resulting $r^{\rm Pugh}$ and $r^{\rm Pett}$ values are highly reliable for our high-throughput screening.

\FloatBarrier
\section{Correlation between \Tc~and  Pettifor's ratio}
\label{sec:sections3}

\begin{figure}[!htb]
    \centering
    \includegraphics[width=0.99\columnwidth]{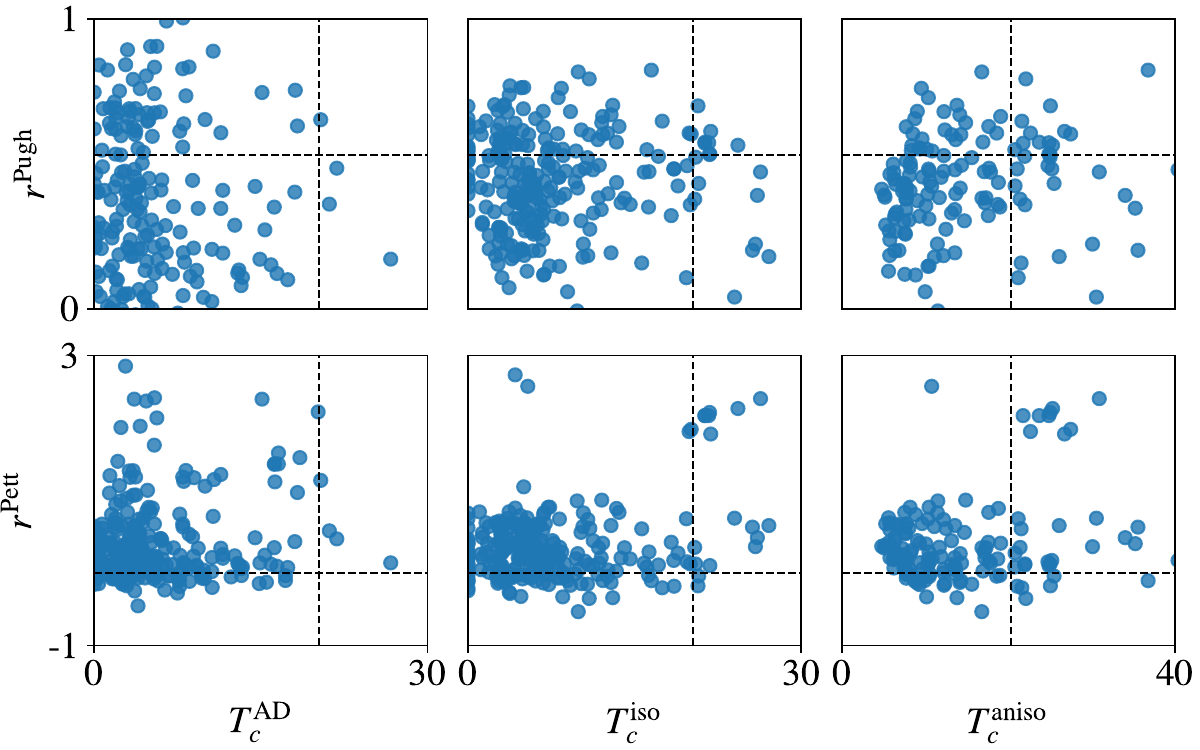}
    \caption{
        Correlation between superconducting transition temperatures and the ductility criteria $r^{\rm Pugh}$ and $r^{\rm Pett}$.
        The ductility thresholds of $r^{\rm Pugh}=0.57$ and $r^{\rm Pett}=0$, along with a transition temperature $T_c=20$~K, are marked by dashed lines.
    }
    \label{fig:figs5}
\end{figure}

In Fig.~\ref{fig:figs5}, we examine the relationship between the superconducting transition temperatures (Allen-Dynes $T_c^{\rm AD}$, isotropic $T_c^{\rm iso}$, and anisotropic $T_c^{\rm aniso}$) and the ductility criteria $r^{\rm Pugh}$ and $r^{\rm Pett}$.

\FloatBarrier
\section{Stacking fault energy}
\label{sec:sections4}

Within the \textsc{aiida-mechanical} package, we implemented the \textsc{GSFEWorkChain} and \textsc{GSFERelaxWorkChain} workflows for generalized stacking fault energy calculations.
These workflows rely on a tilted-cell algorithm, which builds a supercell normal to the slip plane and generates faulted configurations by applying relative displacements along the slip direction.
The process depends heavily on \textsc{aiida-quantumespresso} and executes the following sub-processes sequentially:

\begin{enumerate}
 \item Run \textsc{PwRelaxWorkChain}~(vc-relax) on the pristine unit cell.
 \item Generate the conventional cell structure and the corresponding faulted structures.
 \item Run \textsc{PwBaseWorkChain}~(SCF) on the conventional cell structure generated in step 2.
 \item For each slip direction, run \textsc{PwRelaxWorkChain} (only the spacings between layers, together with the thickness of supercell are relaxed) or \textsc{PwBaseWorkChain} (for unrelaxed calculations) for the first faulted structure and record the Cartesian coordinate of the $\mathbf{k}$ points.
 \item For each of the other candidate faulted structures, run either \textsc{PwRelaxWorkChain} (only the spacings between layers, together with the thickness of supercell are relaxed) or \textsc{PwBaseWorkChain} (for unrelaxed calculations) using exactly the same Cartesian coordinate of the $\mathbf{k}$ points from step 4.
 \item Collect the energies to construct the stacking fault energy curve.
\end{enumerate}

To test the convergence of the \GSFE~curve, we use the \textsc{GSFEWorkChain} (executing only the \textsc{PwBaseWorkChain} in step 4).
Subsequently, we perform the \textsc{GSFERelaxWorkChain} (executing the \textsc{PwRelaxWorkChain} in step 4) for comparison.
The evaluation of Rice's ratio in the main text uses the constrained out-of-plane relaxed results from the \textsc{GSFERelaxWorkChain}.
For A$_1$-type (face-centered cubic, FCC) metals, slip typically occurs on the \{111\} plane.
Along the [111] direction, the FCC lattice exhibits an ABC three-layer stacking sequence.
We denote the perfect stacking sequence as `ABC'.
Aluminium provides a well-studied example of the three characteristic fault configurations.~\cite{Devlin1981}:

\begin{table*}[htbp]
\caption{
    Generalized stacking fault energies for simple FCC metals at the (111)[$11\bar{2}$] slip system with and without relaxation.
    $\gamma^{\rm USFE}$ denotes the unstable stacking fault energy;
    $\gamma^{\rm ISFE}$ and $\gamma^{\rm ESFE}$ represent the intrinsic and extrinsic stacking fault energies respectively; 
    $\gamma^{\rm UTE}$ is the unstable twinning energy.
    }

\setlength{\tabcolsep}{12pt}
\begin{tabularx}{\linewidth}{l rrrrrrrr}
\toprule
Material 
& 
\multicolumn{2}{c}{$\gamma^{\rm USFE}$~(mJ/m$^2$)}  
&
\multicolumn{2}{c}{$\gamma^{\rm ISFE}$~(mJ/m$^2$)} 
& 
\multicolumn{2}{c}{$\gamma^{\rm UTE}$~(mJ/m$^2$)} 
& 
\multicolumn{2}{c}{ $\gamma^{\rm ESFE}$~(mJ/m$^2$)} \\
\cmidrule(lr){2-3} \cmidrule(lr){4-5} \cmidrule(lr){6-7} \cmidrule(lr){8-9}
& unrelaxed & relaxed  & unrelaxed & relaxed & unrelaxed & relaxed & unrelaxed & relaxed  \\
\midrule
Ag & 106.15 & 91.89 & 10.07 & 14.45 & 116.46 & 99.78 & 12.79 & 16.65 \\
Al & 183.38 & 166.47 & 118.97 & 117.84 & 226.30 & 210.73 & 120.87 & 133.68 \\
Au & 90.57 & 64.96 & 23.26 & 21.10 & 103.74 & 75.39 & 23.02 & 22.95 \\
Cu & 228.76 & 222.35 & 26.48 & 3.46 & 239.09 & 224.83 & 26.53 & 11.01 \\
Pb & 86.86 & 75.42 & 47.24 & 44.56 & 113.94 & 96.27 & 54.34 & 37.66 \\
Pt & 329.83 & 285.50 & 315.69 & 285.95 & 446.98 & 378.91 & 305.51 & 284.66 \\
\bottomrule
\end{tabularx}
\label{table:tables1}
\end{table*}

\begin{enumerate}
    \item Intrinsic stacking fault~(ISF): 
        equivalent to a missing plane in the ideal stacking sequence. 
        The faulted stacking is $\cdots$ABC$^*$BCABC$\cdots$
    \item Extrinsic stacking fault~(ESF): 
        equivalent to an additional plane inserted into the ideal stacking sequence. 
        The faulted stacking is $\cdots$ABC$^*$BABC$\cdots$
    \item Unstable twinning~(UT): 
        the top half of the crystal mirrors the bottom half against the gliding plane. 
        The stacking is $\cdots$ABC$^*$ACBA$\cdots$.
\end{enumerate}

There is one unstable configuration from the pristine phase to the intrinsic faulted structure defined as \USFE, and one unstable configuration from the intrinsic faulted structure to the extrinsic faulted structure defined as \UTE.
The energy profile of this process can be described using the following function:
\begin{align}
    &\gamma^{\text{GSFE,fcc}}(b) = 
    \sum_{i=1}^{4} \left \{ A_i \sin(2i\pi b) + B_i \left[ \cos(4i\pi b) - 1 \right] \right \} \nonumber \\ 
    &+ \begin{cases} 
    2\gamma^{\text{ISF}} \, b, & 0 \le b \le 0.5 \\[6pt]
    2(\gamma^{\text{ESF}} - \gamma^{\text{ISF}})b + (2\gamma^{\text{ISF}} - \gamma^{\text{ESF}}), & 0.5 < b \le 1.0
\end{cases}.
\end{align}

\begin{figure}[!htb]
    \centering
    \includegraphics[width=0.99\columnwidth]{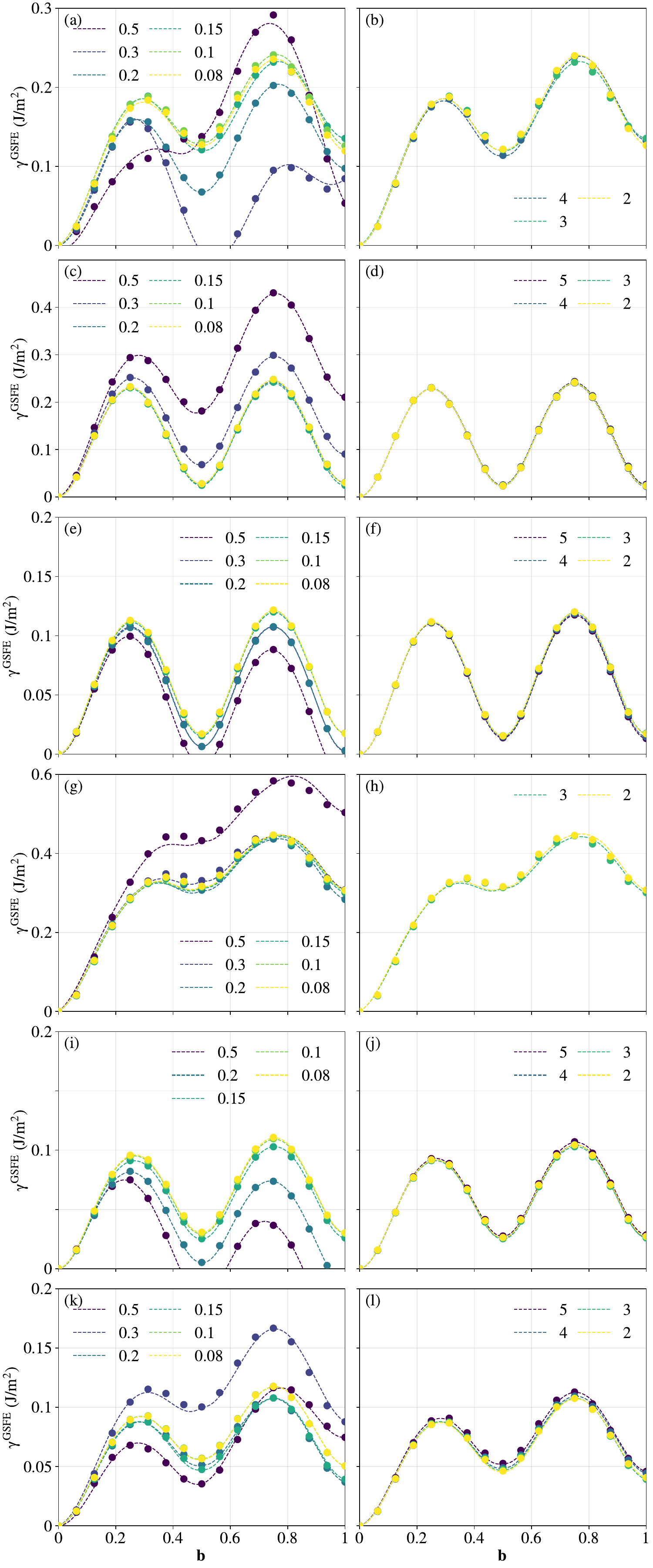}
    \caption{Convergence of the \GSFE~with respect to $\mathbf{k}$-point distance~(${\AA}^{-1}$) (left panels) and supercell size (right panels) for elemental FCC metals: (a,b)~Al, (c,d)~Cu, (e,f)~Ag, (g,h)~Pt, (i,j)~Au, and (k,l)~Pb.}
    \label{fig:figs6}
\end{figure}

\begin{figure}[!htb]
    \centering
    \includegraphics[width=0.99\columnwidth]{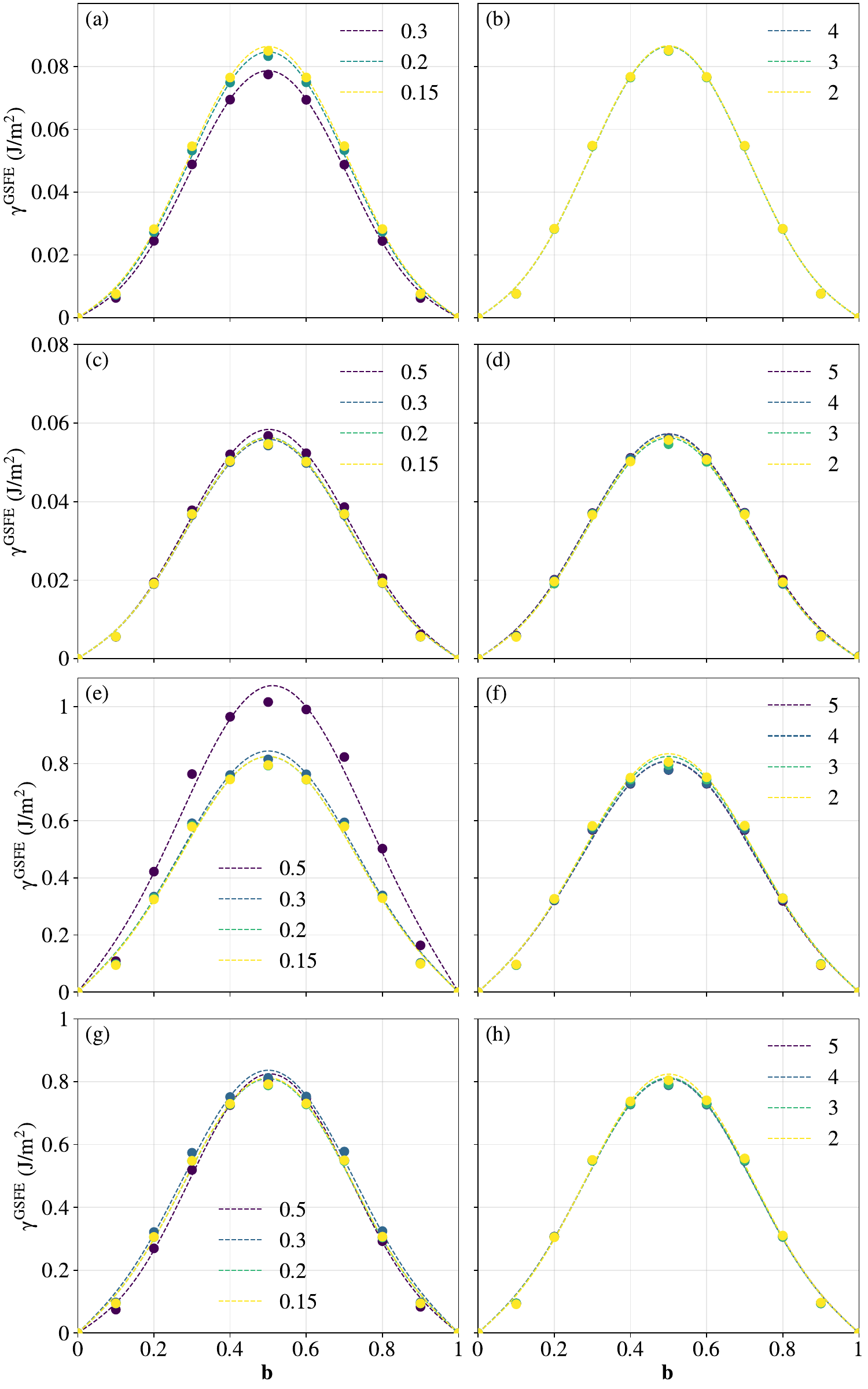}
    \caption{
    Convergence of the \GSFE~with respect to $\mathbf{k}$-point distances~(${\AA}^{-1}$) (left panels) and supercell size (right panels) for elemental BCC metals: (a,b)~Li, (c,d)~Na, (e,f)~V, and (g,h)~Nb.
    }
    \label{fig:figs7}
\end{figure}

We first conduct a convergence test on the $\mathbf{k}$-point distance and supercell size, as shown in Fig.~\ref{fig:figs6}.
For FCC metals, the \GSFE~only converges for $\mathbf{k}$-point distances smaller than $0.15~\AA^{-1}$ and supercells larger than at least two conventional cells.
For materials with a small \ISFE, such as Ag and Au, the predictions can be sensitive to the $\mathbf{k}$-point sampling.
For Ag, we even obtain an unphysical negative \ISFE~using coarse $\mathbf{k}$-point grids.
Using converged parameters, our \GSFE~results are consistent with previous calculations~\cite{Dillamore1965, Bernstein2004, Hartford1998}.
For $\rm{A}_2$-type (body-centered cubic, BCC) metals, the slip occurs at the (1$\bar{1}$0)[111] slip system~\cite{Devlin1981}. 
There is an unstable maximum along the periodic shear path, as noted in Table~\ref{table:tables2}.
The convergence study for BCC metals is shown in Fig.~\ref{fig:figs7}.

\begin{table}[htbp]
\caption{
    \GSFE~for simple BCC metals at the (1$\bar{1}$0)[111] slip system with and without cell relaxation.
    $\gamma^{\rm USFE}$ denotes the unstable stacking fault energy.
    }

\setlength{\tabcolsep}{1.5pt}
\begin{tabularx}{\linewidth}{l@{\hspace{5em}}r@{\hspace{7em}}r}
\toprule
Material 
&
\multicolumn{2}{c}{ $\gamma^{\rm USFE}$~(mJ/m$^2$)} \\
\cmidrule(lr){2-3}
 & unrelaxed & relaxed \\
\midrule
Li & 84.65 & 77.86 \\
Na & 56.36 & 49.96 \\
Nb & 824.87 & 698.61 \\
V & 809.77 & 715.01 \\
\bottomrule
\end{tabularx}
\label{table:tables2}
\end{table}

The $\gamma^{\rm GSFE}$ only converges for $\mathbf{k}$-point distances smaller than $0.2~\AA^{-1}$ and supercells larger than two conventional cells.
For Na, the \USFE~appears converged at a $\mathbf{k}$-point distance of $0.3~\AA^{-1}$.

\begin{figure}[!htb]
    \centering
    \includegraphics[width=0.99\columnwidth]{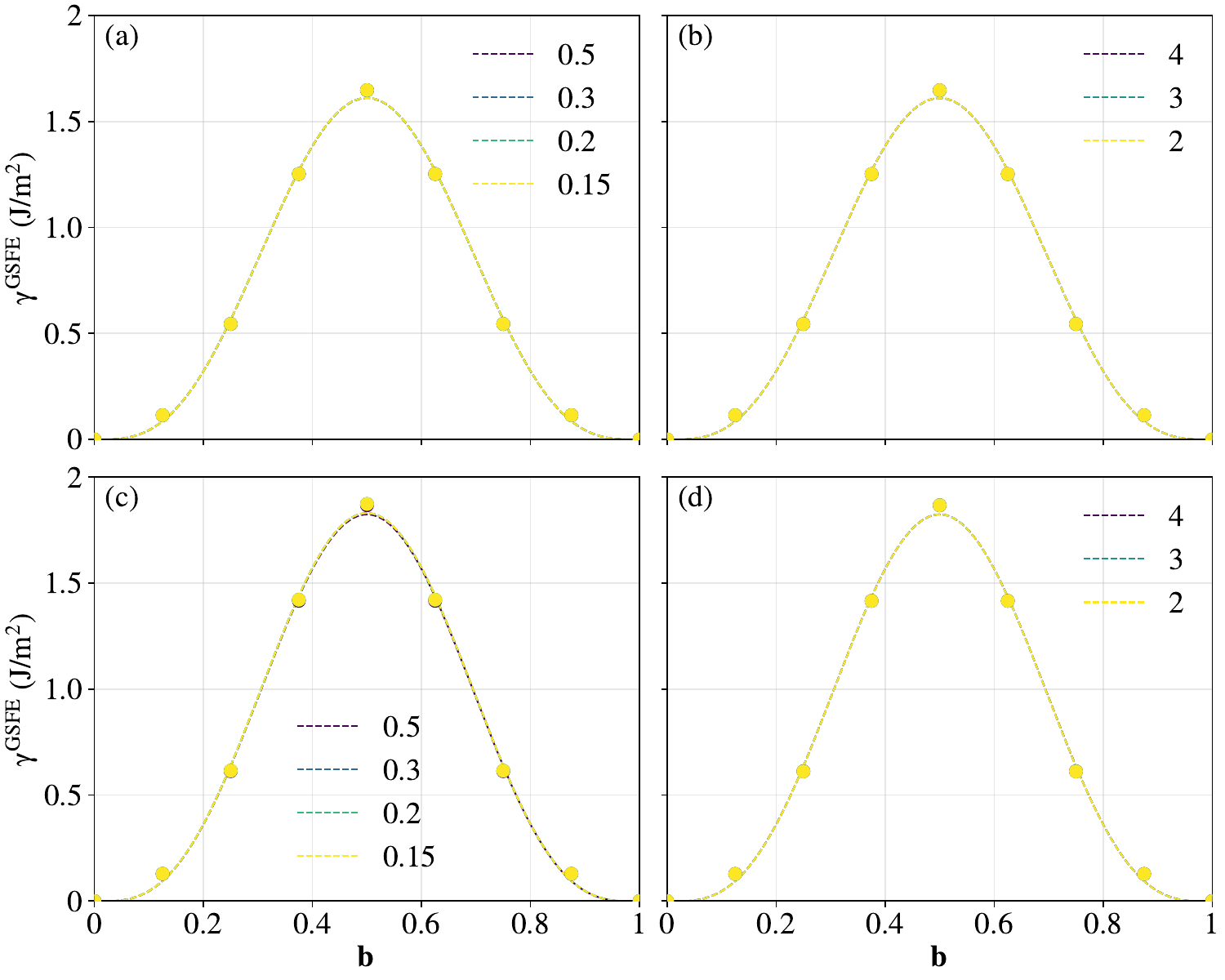}
    \caption{
        Convergence of the \GSFE~with respect to $\mathbf{k}$-point distances~(${\AA}^{-1}$) (left panels) and supercell size (right panels) for $\rm{B}_1$ compounds: (a,c)~NaCl and (b,d)~KCl.
    }
    \label{fig:figs8}
\end{figure}

\begin{figure}[!htb]
    \centering
    \includegraphics[width=0.99\columnwidth]{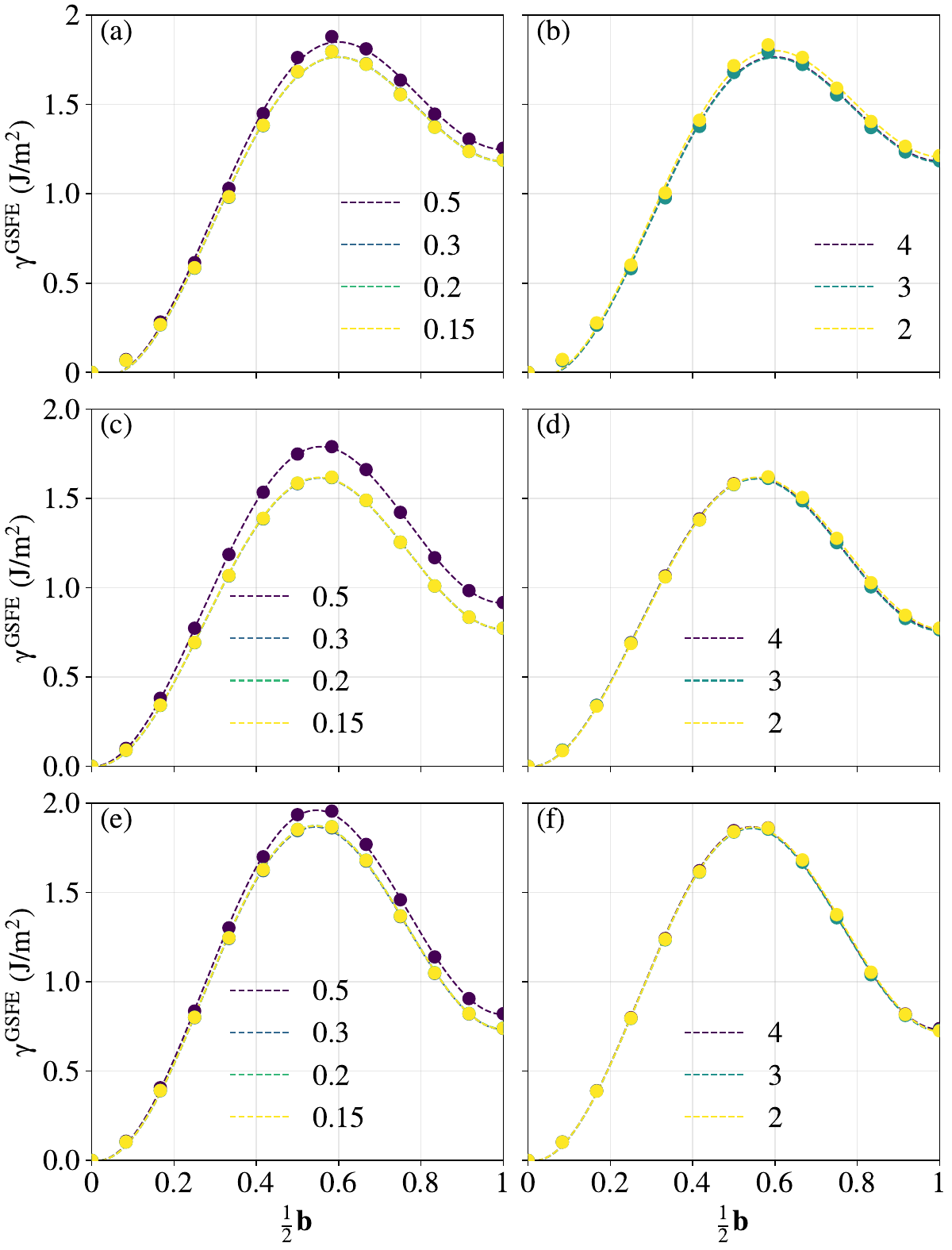}
    \caption{
        Convergence of the \GSFE~with respect to $\mathbf{k}$-point distances~(${\AA}^{-1}$) (left panels) and supercell size (right panels) for: (a,b)~NbCoSb, (c,d)~TiRuSb, and (e,f)~ZrRuSb.}
    \label{fig:figs9}
\end{figure}

For nested FCC structures ($\rm{B}_1$, NaCl-type), the stacking orders and spacings differ from elemental metals.
The most compact slip plane is (1$\bar{1}$0) instead of (111) as in elemental FCC metals.
There is only one unstable stacking fault along this slip system.
As shown by the convergence study in Fig.~\ref{fig:figs8}, the \USFE~is well-converged at a $\mathbf{k}$-point distance of $0.5~\AA^{-1}$ using a supercell larger than two conventional cells.
The convergence tests for the $\rm{C1}_{b}$ crystals (half-Heusler) \ce{NbCoSb}, \ce{TiRuSb}, and \ce{ZrRuSb} on the (1$\bar{1}$0) plane along the [110], [001], and [112] directions are shown in Fig.~\ref{fig:figs9}.

\begin{table}[htbp]
\caption{
    \GSFE~for rock-salt crystals at the (1$\bar{1}$0)[110] slip system with and without relaxation.
    $\gamma^{\rm USFE}$ denotes the unstable stacking fault energy.
    }
\begin{tabularx}{\linewidth}{l@{\hspace{5em}}r@{\hspace{7em}}r}
\toprule
Material 
&
\multicolumn{2}{c}{ $\gamma^{\rm USFE}$~(mJ/m$^2$)} \\
\cmidrule(lr){2-3}
& unrelaxed & relaxed \\
\midrule
KCl & 1204.94 & 154.59 \\
NaCl & 1362.23 & 177.34 \\
\bottomrule
\end{tabularx}
\label{table:tables3}
\end{table}

\begin{figure}[!htb]
    \centering
    \includegraphics[width=0.99\columnwidth]{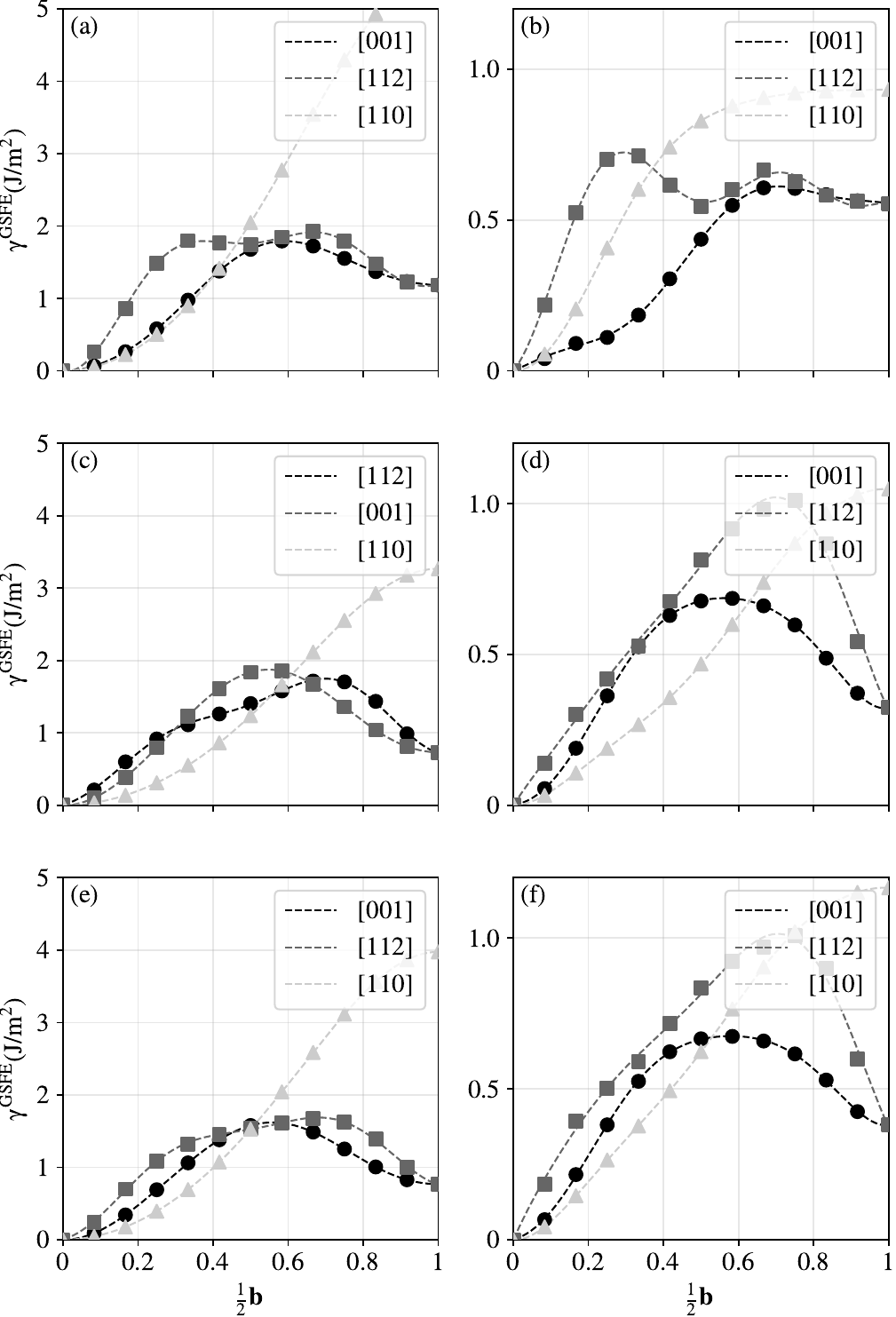}
    \caption{
        \GSFE~along the [110], [001], and [112] directions without relaxation~(the left panels) and with relaxation~(the right panels) for (a, b)~NbCoSb, (c, d)~TiRuSb, and (e, f)~ZrRuSb.
    }
    \label{fig:figs10}
\end{figure}

$\rm{C1}_{b}$ crystals are triply nested FCC structures.
Unlike $\rm{A}_1$ metals, these compounds behave more like $\rm{B}_1$ materials, showing weak sensitivity to the $\mathbf{k}$-point density and supercell size.
The $\gamma^{\rm GSFE}$ curve converges rapidly at a $\mathbf{k}$-point distance of $0.3~{\AA}^{-1}$ with a supercell of two conventional layers.
The easiest slip direction identified in the convergence tests is [001].
Incorporating structural relaxation along the $z$-axis significantly reduces the $\gamma^{\rm GSFE}$, as depicted in Fig.~\ref{fig:figs10}.
The \USFE~values for both unrelaxed and relaxed calculations are summarized in Table~\ref{table:tables4}, illustrating this significant reduction upon relaxation.

\begin{table}[htbp]
\caption{
    \GSFE~for C1$_b$ compounds at the (1$\bar{1}$0) slip plane with and without cell relaxation.
    $\gamma^{\rm USFE}$ denotes the unstable stacking fault energy.
    }
\begin{tabularx}{\linewidth}{l@{\hspace{5em}}r@{\hspace{7em}}r}
\toprule
Material 
&
\multicolumn{2}{c}{ $\gamma^{\rm USFE}$~(mJ/m$^2$)} \\
\cmidrule(lr){2-3}
& unrelaxed & relaxed \\
\midrule
NbCoSb & 1765.81 & 617.47 \\
TiRuSb & 1866.79 & 701.26 \\
ZrRuSb & 1614.46 & 686.99 \\
\bottomrule
\end{tabularx}
\label{table:tables4}
\end{table}

\begin{figure}[!htb]
    \centering
    \includegraphics[width=0.99\columnwidth]{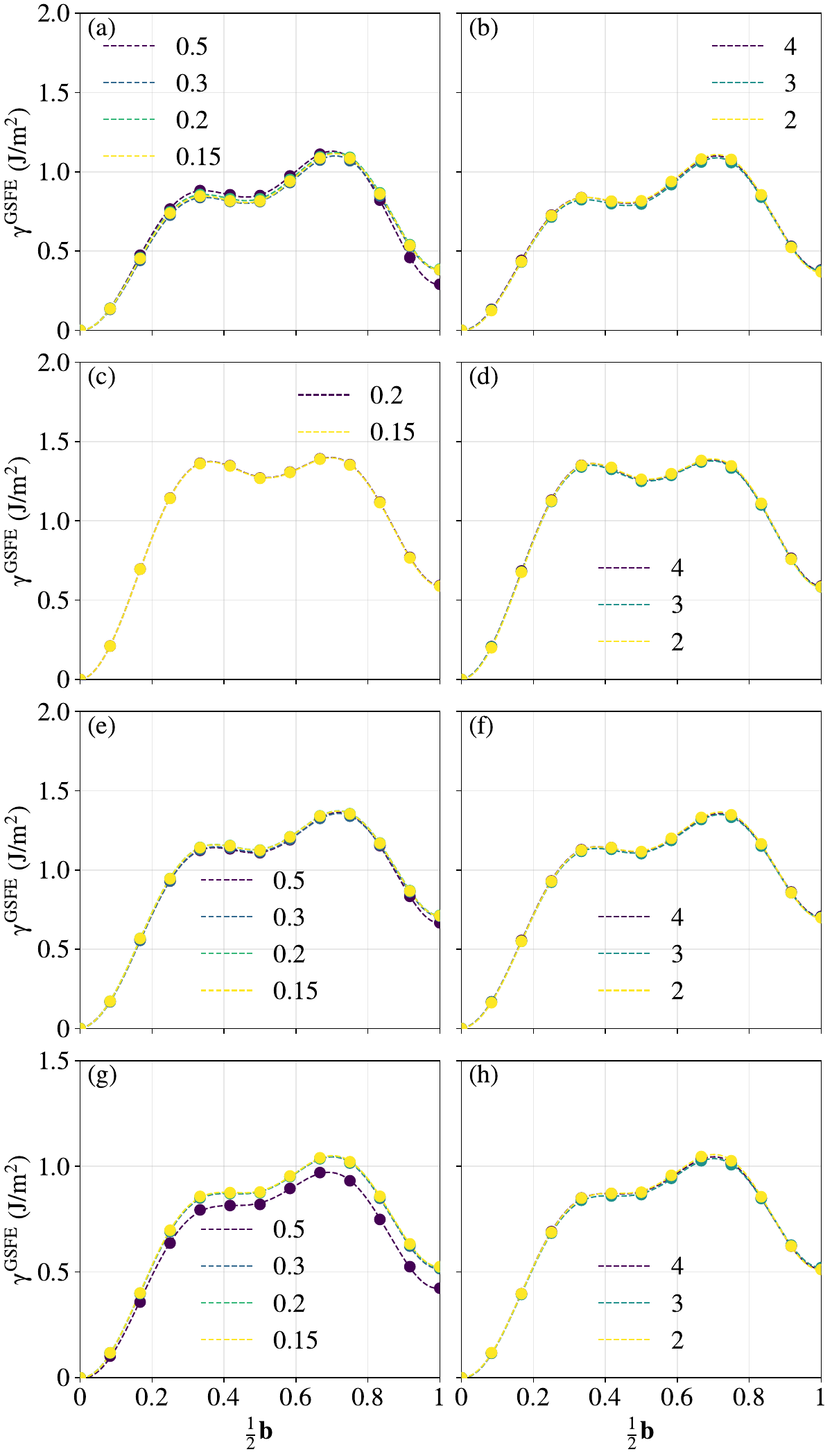}
    \caption{
        Convergence of the \GSFE~with respect to $\mathbf{k}$-point distances~(${\AA}^{-1}$) (left panels) and supercell size (right panels) for: (a,b)~\ce{HfPd2Al}, (c,d)~\ce{ZrNi2Al}, (e,f)~\ce{ZrNi2Ga}, and (g,h)~\ce{YPd2Sn}.
        }
    \label{fig:figs11}
\end{figure}

\begin{figure}[!htb]
    \centering
    \includegraphics[width=\columnwidth]{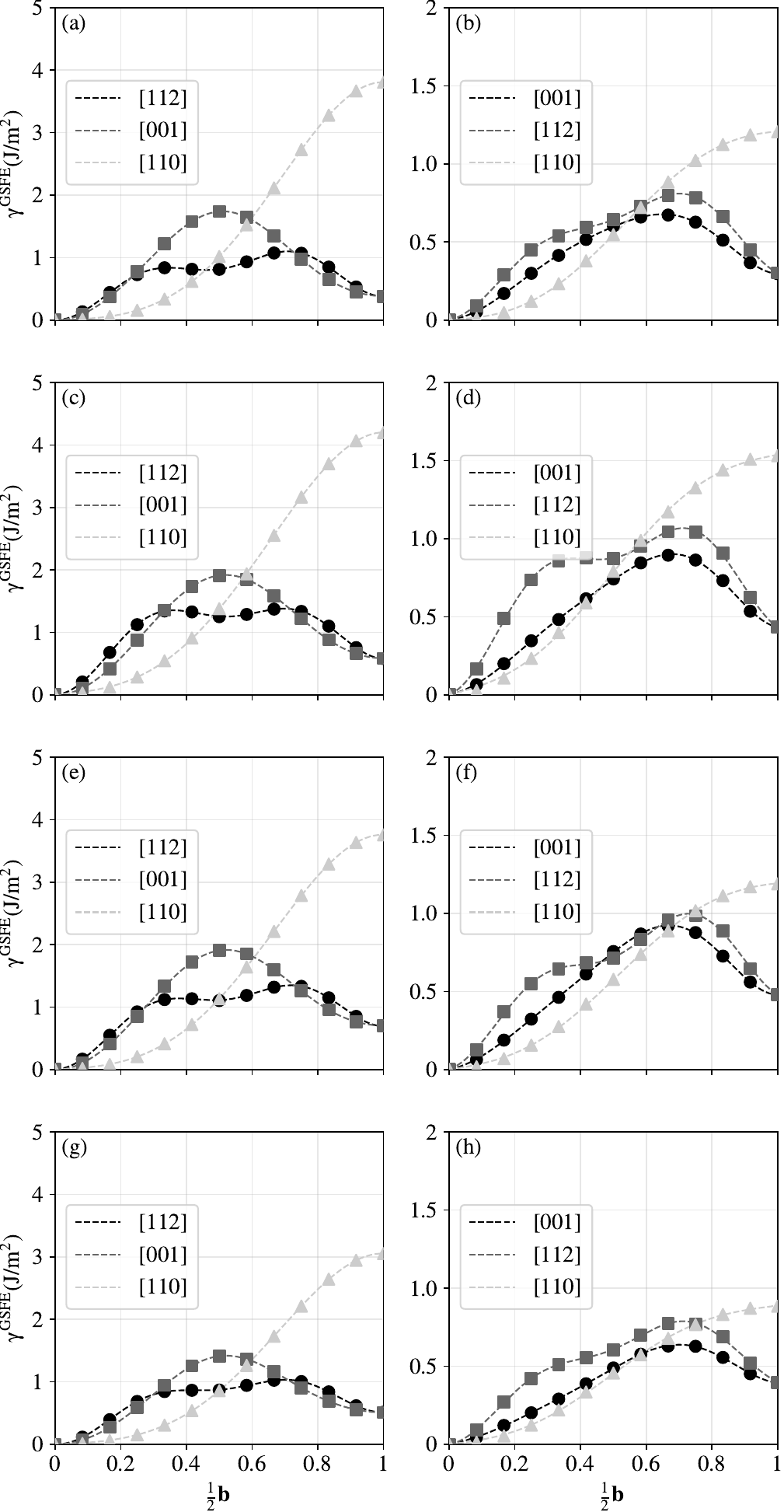}
    \caption{
        \GSFE~along the [110], [001], and [112] directions without relaxation~(the left panels) and with relaxation~(the right panels) for (a, b)~\ce{HfPd2Al}, (c, d)~\ce{ZrNi2Al}, (e, f)~\ce{ZrNi2Ga}, and (g, h)~\ce{YPd2Sn}.
    }
    \label{fig:figs12}
\end{figure}

Convergence tests were also performed for four L2$_1$-type Heusler compounds: \ce{HfPd2Al}, \ce{ZrNi2Al}, \ce{ZrNi2Ga}, and \ce{YPd2Sn}.
The $\gamma^{\rm GSFE}$ curve converges at a $\mathbf{k}$-point distance of $0.3~{\AA}^{-1}$ with a supercell of two conventional layers.
In the unrelaxed state, we observe the barrier hierarchy $\gamma^{\rm USFE, [110]} > \gamma^{\rm USFE, [001]} > \gamma^{\rm USFE, [112]}$.
In Fig.~\ref{fig:figs11}, we show the convergence tests specifically along the [001] direction.
We then computed the $\gamma^{\rm GSFE}$ with relaxation along the $z$-axis, as shown in Fig.~\ref{fig:figs12}.
With the relaxation of the cell and atomic positions, the easiest slip direction changes, resulting in the hierarchy $\gamma^{\rm USFE, [110]} > \gamma^{\rm USFE, [112]} > \gamma^{\rm USFE, [001]}$.
The \USFE~values for unrelaxed and relaxed calculations are summarized in Table~\ref{table:tables5}, again demonstrating a significant reduction in \USFE~upon relaxation.

\begin{table}[htbp]
\caption{
    \GSFE~of L2$_1$ materials at the (1$\bar{1}$0)[001] and (1$\bar{1}$0)[112] slip systems with and without relaxation.
    $\gamma^{\rm USFE}$ denotes the unstable stacking fault energy.
    }
\begin{tabularx}{\linewidth}{l r@{\hspace{2.2em}} r @{\hspace{2.2em}} r @{\hspace{2.2em}} r}
\toprule
Material 
&
\multicolumn{4}{c}{ $\gamma^{\rm USFE}$~(mJ/m$^2$)} \\
\cmidrule(lr){2-5}
& \multicolumn{2}{c}{unrelaxed} & \multicolumn{2}{c}{relaxed} \\
\cmidrule(lr){2-3} \cmidrule(lr){4-5} 
& [001] & [112]  & [001] & [112] \\
\midrule
HfPd$_2$Al & 1762.79 & 1121.36 & 674.26 & 767.04 \\
ZrNi$_2$Al & 1928.70 & 1399.60 & 887.42 & 998.55 \\
ZrNi$_2$Ga & 1930.09 & 1373.23 & 910.81 & 940.69 \\
YPd$_2$Sn & 1423.87 & 1047.06 & 630.80 & 752.08 \\
\bottomrule
\end{tabularx}
\label{table:tables5}
\end{table}

\FloatBarrier
\section{Surface energy}
\label{sec:sections5}

Within the \textsc{aiida-mechanical} package, we also implement the \textsc{SurfaceEnergyWorkChain} workflow for surface energy calculations, which automatically tests for convergence with respect to the vacuum thickness according to the following procedures:

\begin{enumerate}
 \item Run \textsc{PwRelaxWorkChain}~(vc-relax) on the pristine unit cell.
 \item Generate the conventional cell structure and cleaved slab structures for a range of vacuum ratios using the relaxed structure from step 1.
 \item Run \textsc{PwBaseWorkChain}~(SCF) on the conventional cell structure generated in step 2.
 \item For each candidate cleaved structure, run \textsc{PwBaseWorkChain}~(SCF).
 \item Collect the total energies and compute the surface energies.
\end{enumerate}

\begin{figure}[!htb]
    \centering
    \includegraphics[width=0.99\columnwidth]{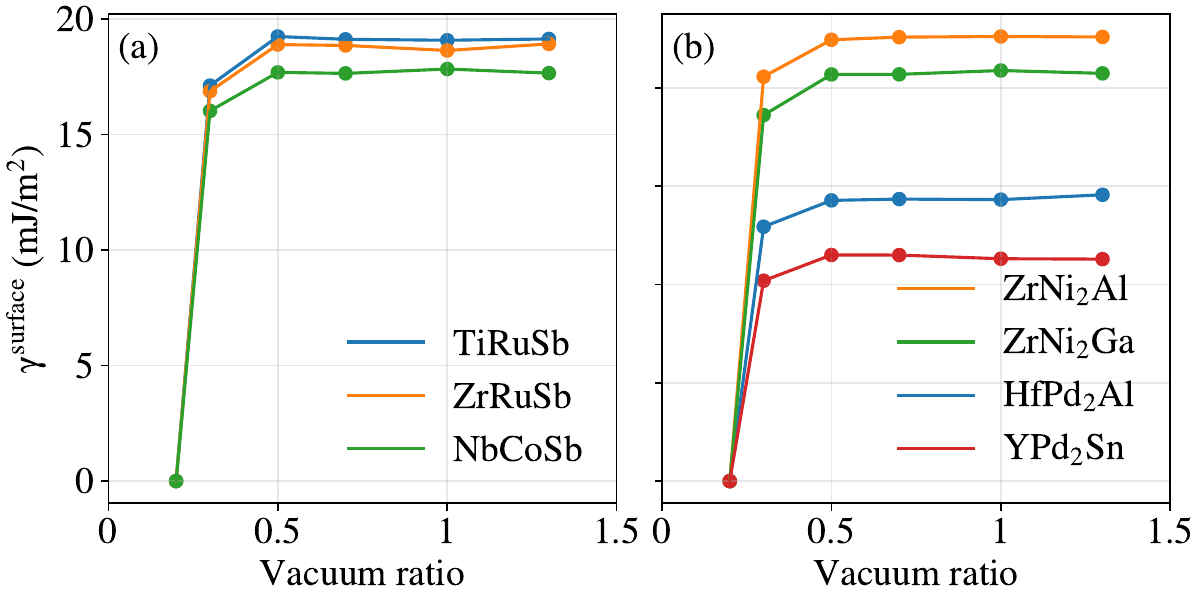}
    \caption{
        Convergence of the surface energy as a function of the relative vacuum spacing for 
        (a)~C1$_b$ (half-Heusler) materials \ce{NbCoSb}, \ce{TiRuSb}, and \ce{ZrRuSb} and (b)~L2$_1$ (Heusler) materials \ce{HfPd2Al}, \ce{ZrNi2Al}, \ce{ZrNi2Ga}, and \ce{YPd2Sn}.
        }
    \label{fig:figs13}
\end{figure}

We converged the calculated surface energies with respect to the vacuum spacing for both the $\rm{C1}_{b}$ and $\rm{L2}_{1}$ slabs.
We define the \emph{relative vacuum} as the ratio between the vacuum thickness and the slab thickness.
In Fig.~\ref{fig:figs13}, we find that for relative vacuum values above 0.5, the surface energies fluctuate by less than 0.01~J/m$^2$.
Therefore, in this work, we use a relative vacuum of 1.0, ensuring that the reported surface energies are converged.

\bibliographystyle{elsarticle-num}

\bibliography{Bibliography}

\end{document}